# Quantifying Policy Responses to a Global Emergency:
# Insights from the COVID-19 Pandemic

Jian Gao[1,2,3¶], Yian Yin[1,2,4¶], Benjamin F. Jones[1,2,3,5*], Dashun Wang[1,2,3,4*]

[1]Center for Science of Science and Innovation, Northwestern University, Evanston IL

[2]Northwestern Institute on Complex Systems, Northwestern University, Evanston IL

[3]Kellogg School of Management, Northwestern University, Evanston IL

[4]McCormick School of Engineering, Northwestern University, Evanston IL

[5]National Bureau of Economic Research, Cambridge MA

[¶]These authors contributed equally to this work.

*Correspondence to: bjones@kellogg.northwestern.edu, dashun.wang@northwestern.edu.

**Public policy must confront emergencies that evolve in real time and in uncertain directions [1-6], yet little is known about the nature of policy response. Here we take the coronavirus pandemic as a global and extraordinarily consequential case [7-12], and study the global policy response by analyzing a novel dataset recording policy documents published by government agencies, think tanks, and intergovernmental organizations (IGOs) across 114 countries (37,725 policy documents from January 2nd through May 26th 2020). Our analyses reveal four primary findings. (1) Global policy attention to COVID-19 follows a remarkably similar trajectory as the total confirmed cases of COVID-19, yet with evolving policy focus from public health to broader social issues. (2) The COVID-19 policy frontier disproportionately draws on the latest, peer-reviewed, and high-impact scientific insights. Moreover, policy documents that cite science appear especially impactful within the policy domain. (3) The global policy frontier is primarily interconnected through IGOs, such as the World Health Organization, which produce policy documents that are central to the COVID-19 policy network and draw especially strongly on scientific literature. Removing IGOs' contributions fundamentally alters the global policy landscape, with the policy citation network among government agencies increasingly fragmented into many isolated clusters. (4) Countries exhibit highly heterogeneous policy attention to COVID-19. Most strikingly, a country's early policy attention to COVID-19 shows a surprising degree of predictability for the country's subsequent deaths. Overall, these results uncover fundamental patterns of**

**policy interactions and, given the consequential nature of emergent threats and the paucity of quantitative approaches to understand them, open up novel dimensions for assessing and effectively coordinating global and local responses to COVID-19 and beyond.**

**The global policy response to the COVID-19 pandemic**

Human society is often challenged by emergencies, from natural disasters [1-3] and pandemics [7, 8, 13, 14] to human conflicts [15, 16] and other emergent crises [3, 17, 18]. Despite the consequential nature of public response [6-12], governments often struggle to manage these emergent problems, potentially due to various reasons ranging from the inherent difficulties in altering rules and allocating attention and resources [19, 20] to uncertainties about the evolving nature of the problems [19-21].The COVID-19 pandemic provides a high-stakes case for examining the nature of policy response [5-12, 20]. Indeed, the coronavirus pandemic is not only a public health crisis but also creates enormous challenges to social, economic, and political systems around the world [19, 20, 22]. At the same time, evolving knowledge about the nature of the disease [23, 24], as well as evolving knowledge about containment policies and social implications [5, 7-11, 19, 20, 22], presents a dynamic and uncertain policy environment.

To quantify the global policy response to the COVID-19 pandemic, we analyze a novel dataset of policy documents (37,725 documents published from January 2nd to May 26th, 2020). The dataset categorizes the source, topics, publication date of each policy document, and further includes each document's cited references. The institutions include government agencies, think tanks, and intergovernmental organizations (IGOs). The coverage is worldwide, with policy documents from 114 countries, and includes all major economies and large population centers, with a notable exception of mainland China, overall covering ~66.3% of the world population, ~79.3% of total Gross Domestic Product (GDP), and ~95.6% of confirmed deaths worldwide due to COVID-19 (as of May 30th, see Fig. S1 and SI S1.1 for further description). Within this corpus, we identify COVID-19 related policy documents through keyword filtering (7,730 documents in total, see SI S1.1), allowing us to compare COVID-19 policy documents with all other policy documents published in 2020. For each policy document, we further analyze all the references cited therein, allowing us to link policy documents to scientific papers and other policy documents they reference. These linkages can inform how different institutions interact in policy making and also illuminate the nature of the scientific knowledge upon which these policies draw.

Overall, the total number of policy documents recorded in our database grows roughly linearly with time, averaging approximately 280 new policy documents per day (Fig. 1A). By contrast, the number of COVID-19 related policy documents features highly non-linear growth, approximated by an exponential pattern (Fig. 1B and inset), showing rapid shifts in global policy attention over the past few months. To understand how policy attention tracks the coronavirus outbreak, we measure the share of policy attention to COVID-19 by calculating the proportion of COVID-19 related policy documents among all policy documents published on a given day. We then compare this share with how the pandemic unfolded globally, as traced by the Johns Hopkins COVID-19 tracking map [23]. We find that the share of COVID-19 policy attention shows a remarkably similar trajectory as the total confirmed cases of COVID-19 over time (Fig. 1C, red vs blue). We observe the same pattern when using total confirmed deaths (Fig. S2). Together, these results show that the share of policy attention to COVID-19 closely corresponds to the global unfolding of the pandemic (see SI S3.1 for more detail).

To inform how the COVID-19 policy attention may reflect policy-making institutions' priorities, we further examine the content of the COVID-19 related policy documents, breaking them down by field (Fig. 1D) and topic (Fig. 1E). For both analyses, we observed clear shifts in policy attention related to the pandemic, with policy priorities shifting from issues in public health to economy and society. Indeed, we first classify COVID-19 related policy documents by field, grouping them into three major field categories ("science & health", "economy & labour", and "society & others", see SI S2.1 for more detail). Plotting the share of COVID-19 policy documents across these field categories (Fig. 1D), we find that in the early stage of the outbreak (January and February 2020), about 90% of COVID-19 policies belong to the health and science category, showing a clear, initial focus on public health issues. The policy priorities show a visible shift, however, since early March 2020, with a notable rise in policy attention to issues around the economy and society, suggesting a growing policy balance between health and socio-economic implications of the pandemic (see Fig. S3B for results at the individual field level).

We further break down these policy documents at the narrower topic level, plotting dynamics related to COVID-19 for the top 10 topics (Fig. 1E). We find again an initial focus on health-related topics (in blue), followed by a clear increase over time in the share of topics related to economy and society (in red). Indeed, Fig. 1F-G respectively show the word cloud of all topics pre- and post-WHO's pandemic declaration, which illustrates a stark difference in policy priorities across the two periods, shifting sharply from public health to socioeconomic issues (see Fig. S3A for topic changes by month). Furthermore, the shifts documented in Figs. 1D-G are observed in COVID-19 policy documents only, as we repeated all our analyses for other policy documents published in the same period, finding that their shares by field or topic are relatively stable over time (Fig. S4).

**The role of science in the COVID-19 policy response**

Overall, Fig. 1 demonstrates a rapidly evolving global policy frontier that closely corresponds to the evolution of the pandemic. Yet at the same time the scientific research related to COVID-19 also evolves rapidly, as exemplified by the strong response from the global research enterprise [5, 20]. Indeed, as of May 30th 2020, over 40,000 papers on coronavirus research have been published in 2020, with preprint servers playing an outsize role in disseminating the latest science [22, 25]. The parallel advancement of the policy and scientific frontier in response to the pandemic raises a crucial question: is our policy understanding of COVID-19 closely linked with scientific understanding, or largely separated from the ivory tower? To answer this question, here we trace all scientific papers cited by policy documents in our corpus and match them to large-scale publication and citation databases (SI S1.2), offering a unique opportunity to quantify the role of science in the global policy response to COVID-19.

We find that COVID-19 policy documents rapidly evolve with the scientific frontier. First, Fig. 2A shows that the probability for these policy documents to cite at least one scientific paper fluctuates until WHO's pandemic declaration, and features a steady increase afterward. Fig. 2B plots the age distribution of scientific evidence referenced by COVID-19 policy documents, showing that COVID-19 policies are disproportionately centered on the latest scientific frontier. Indeed, out of all scientific references drawn upon by COVID-19 policy documents, 19.9% of the scientific papers were published in 2020. This rate of utilizing the most recent scientific knowledge is over ten times larger than seen for other policy documents. Not surprisingly, we find that the latest scientific evidence cited by COVID-19 policies is primarily related to research on coronavirus (88.4%). However, as policy attention shifts gradually from health to economy over the course of the pandemic (Fig. 1D-E), we also observe a similar shift in the fields of science that COVID-19 policies cite (Fig. 2C), showing a clear shift from drawing primarily upon the biomedical literature to citing economics, society, and other fields of study (Figs. S3C).

Together, these results suggest that despite the extremely recent development in COVID-19 related scientific research, new scientific work substantially finds its way into policy documents, prompting us to next examine the quality of scientific evidence that informs policy. To proceed, we separate COVID-19 related papers into two groups based on whether or not they are referenced by policy documents. We then measure each paper's scientific impact, approximated by the number of citations the paper receives from other scientific papers. We find a dramatic impact difference between the two groups (Fig. 2D): papers referenced in policy documents garner on average 40 times higher citations than those not referenced in policy (average citations: 67.72 vs 1.67). This result shows that the coronavirus research used by policymakers tends to align with what scientists themselves consider important.

Figure 2E further breaks down the policy coverage of COVID-19 related research based on publication venues. We find that different venues differ widely in publication volume, with preprint servers such as medRxiv, bioRxiv, and SSRN publishing an order of magnitude more COVID-19 related papers than peer-reviewed journals. Yet, despite the large amount of papers posted on preprint servers, their impact in policy is rather limited, as these preprint servers show consistently fewer policy citations than average. By contrast, COVID-19 policies disproportionately reference peer-reviewed insights, drawing especially heavily on respected medical journals, both general (e.g., Lancet) and specialized (e.g., Clinical Infectious Diseases). Amid growing concerns over the quality and abundancy of coronavirus research posted on preprint servers [26], these results show that peer-reviewed journals appear to remain an important institution in supplying scientific evidence for policy making.

Overall, the COVID-19 policy frontier appears grounded in extremely recent, peer-reviewed scientific insights, and science directly drawn by this policy frontier is viewed as especially impactful within the scientific community itself. Our next set of analyses present the same pattern but on the other side of the policy-science interface, showing that policy documents that are grounded in the scientific frontier tend to gather substantially more use within the global policy network. Figure 2F separates COVID-19 policy documents by whether they cite any scientific reference or not, and compares the average number of citations from other policy documents. We find that COVID-19 policy documents that cite at least one scientific paper are associated with a significantly higher number of citations from other policies, more than doubling its policy use than those without scientific references. To test if the difference observed in Fig. 2F can be explained by other covariates, we further use a regression model (see Table S1 and SI S3.3 for more detail) to control for policy sources, the total number of scientific references, the date of policy documents, and self-citations, finding our conclusions remain the same ($P < 0.001$). Together, the results in Fig. 2 show that, despite the rapidly evolving nature of the pandemic, the policy and scientific frontier of COVID-19 are closely interlinked, with advances directly along the policy-science interface being notably more impactful within their own domains.

**The role of institutions in the COVID-19 policy response**

Figure 3 considers the varied role of policy-making institutions in the global policy frontier of COVID-19. Analyzing policy documents from three types of institutions (government agencies, think tanks, and IGOs), we find that overall IGOs produce the fewest number of documents (Fig. 3A), yet the policy documents they produce tend to be most closely connected to the scientific frontier (Fig. 3B) and occupy a central position within the global policy network (Fig. 3C-E). We visualize the policy citation network among COVID-19 policy documents in Fig. 3E. Each node is a COVID-19 policy document with at least one citation linkage to other COVID-19 policy documents, and two nodes are connected if one policy cites the other. We color the nodes based on the institution that published the policy (IGOs in blue, think tanks in orange, and governments in green). A node's size is proportional to the number of citations received within this policy network. We also add a node border (in black) for those that draw on scientific papers.

The networks convey three primary insights (Fig. 3C-E). First, although IGOs produced the fewest number of policy documents among the three types of institutions (Fig. 3A), blue nodes are the most visible within the network, indicating IGOs are disproportionately more likely to produce hubs within the policy network (Fig. 3F). Second, hubs in this policy network mostly have black borders, indicating that those that attracted more policy use are also directly grounded in science, consistent with Fig. 2F. Third, while the connectivity among policy documents is relatively sparse (2,994 linkages between 2,098 nodes, $<k_{in}> = <k_{out}> = 1.43$), a high fraction of policy documents coalesce into the same giant weakly connected component (GWCC, 70.3%), illustrating the connectedness within the global policy frontier.

To quantify the relative locations of policy documents within the network, we use the $k$-core algorithm to calculate the coreness of each node (Fig. 3F inset, SI S3.2). A popular concept to measure the core-periphery structure of complex networks, $k$-core is a maximal subgraph of the original network, all of which are connected to at least $k$ other nodes in the group [27]. Figure 3F inset compares the distribution of coreness for policy documents produced by the three institutions, showing policy documents produced by IGOs systematically occupy central positions within the network, characterized by a high $k$-core score. More importantly, these IGO documents appear critical for the global connectivity of this policy network. In Fig. 3G we randomly remove a fixed number of nodes from certain types and measure the remaining GWCC size [28, 29], finding IGO documents carry disproportionate importance in connecting the network. Indeed, as Fig. 3E shows, removing IGOs' policy documents drastically alters the structure of the policy network. While IGOs only account for 31.0% of the nodes (Fig. 3C), removing them shrinks the GWCC from 70.3% to 18.0 % (Fig. 3DG). Contrasting IGOs with other institutions, we find that the effects of removing policy documents by think tanks or government agencies are not nearly as significant (Fig. 3D-EG). Lastly, the fraction of the GWCC decreases to merely 5.77% if we further remove all think tank policy documents, keeping only those by government agencies (Fig. 3E), highlighting that policies by individual countries rarely connect directly across geopolitical boundaries.

**The effectiveness of local COVID-19 policy responses**

Taken together, Fig. 3 shows that the overall connectedness in the COVID-19 policy frontier is mainly mediated by IGOs, the absence of which alters the global policy landscape. And among IGOs, the WHO plays a dominant role across all 55 organizations recorded in our data (Fig. 3H). While these results highlight the critical role of IGOs and especially the WHO in policy responses to COVID-19 [30, 31], they also suggest that the apparent connectivity in the global COVID-19 policy network hides a high degree of isolation at the individual governmental level. Indeed, if we only consider policy responses by government agencies, individual countries are fragmented into small isolated clusters, as their COVID-19 policies primarily interact within their geopolitical boundaries. This suggests that individual countries may differ in their policy response to COVID-19 [6, 32], prompting us to examine whether country-level variations in COVID-19 policy attention may predict country-specific realities of the pandemic.

Therefore, in our final analyses, we consider the COVID-19 policy response at the individual country level, and explore its correlation with the country's effectiveness in containing the pandemic (Fig. 4). For each country in our database, we mark the calendar date when three COVID-19 deaths were first recorded within the same day [33]. We then measure the country's total deaths from COVID-19 over the following 30 days, which approximates the country's effectiveness in containing the disease. We further calculate the share of policy attention to COVID-19 in that country during the prior 30 days, i.e., in the 30 days before three daily deaths were first recorded in that country. Plotting the results in Fig. 4A, we find a strong negative correlation ($n = 59$, $P < 0.00001$), showing that countries with a greater share of policy documents addressing COVID-19 experienced substantially fewer subsequent deaths.

Several additional analyses help further assess this result. Specifically, countries with larger populations may naturally experience greater deaths. Countries with greater per-capita income may also be more susceptible (e.g., through higher influxes of international travel). Finally, countries that experienced later onset of the outbreak may have more time to prepare and learn from others. Table S2 considers various regression models to control for these additional features. We find robust evidence that greater population predicts greater death and that later onset of the disease is associated with less ensuing death, which is consistent with global learning about how to handle the pandemic. At the same time, the explanatory power of a country's early policy attention remains robust. Figure 4B shows the ongoing predictive role of a country's prior policy attention, net of other controls ($P = 0.043$). Figure 4C shows the collective power of prior policy attention and the other controls in predicting subsequent deaths, demonstrating the substantial predictive capacity of this simple model.

We further test the robustness of these findings across several additional dimensions. We restrict the analysis to countries with higher volumes of policy documents recorded in our database (Tables S3-S4), examine different daily death thresholds to mark the onset of the disease for each country (Fig. S7) and consider alternative functional forms (Tables S2-S4). We also adopt alternative data sources for the COVID-19 tracking data, as provided by the European Centre for Disease Prevention and Control (Fig. S5D), alter the measures of a country's effectiveness in containing the pandemic by calculating total confirmed cases instead of deaths (Fig. S5A), examine per-capita death rates for each country (Fig. S5B), and restrict the analysis to countries with a minimum number of COVID-19 policy documents (Fig. S6). The findings are broadly robust to these many alternatives. Finally, we use an alternative policy attention measure – the stringency index provided by the Oxford COVID-19 Government Response Tracker (OxCGRT) [6] – which is developed independently of the Overton policy document database we studied. This alternative source shows a similar relationship (Table S5), where greater policy stringency predicts fewer subsequent deaths, and further suggests the robustness of our result.

Overall the results suggest a country's early policy attention to COVID-19 predicts its containment of the outbreak. The patterns presented in Fig. 4 are correlations and do not identify causal mechanisms. Indeed, there are a large number of social, cultural, economic, and political factors that may influence a country's response to COVID-19. Yet, despite these many potential influences and complex factors, a country's early policy attention provides a simple yet powerful proxy that closely predicts a country's effectiveness in containing the pandemic, offering a new dimension in measuring heterogeneous policy responses and predicting differences in outcomes.

**Concluding remarks**

While the data used in this paper represents among the largest collection of policy documents worldwide, there are limitations of the data that should be considered when interpreting the results (SI S1.4, S3.4-S3.10). First, policy documents from mainland China, where the COVID-19 originated, are notably missing from the data. Second, many policy documents are written in the native language of a country, requiring an additional machine translation step to integrate the corpus, which was performed at the data source. It is important to keep this language effect in mind, which may generate differential data quality for non-English speaking countries. It is worth noting that potential sampling biases in our data are likely to be conservative. Indeed, we repeated our analyses in Fig. 4 for English-speaking countries only (SI S3.7), finding that the relationship further strengthens (Fig. S5C). Lastly, the data focus on policy responses at the country level, while across regions within each country there can be important local differences in response [12, 34, 35], which are not captured by our data.

The data and methodologies presented in our paper are not limited to analyzing COVID-19 policy, highlighting several fruitful future avenues to quantitatively investigate policies, including their rich connections to other policies as well as to the scientific literature. Future work may extend our analyses to other large-scale emergencies or historical events. Equally fruitful is the potential to quantitatively analyze patterns of policy interactions under normal circumstances, which may uncover new signals for relevant social economic indicators. Lastly, the linkage between policy documents and scientific papers opens up new possibilities for understanding how policy and science interact, which would not only broaden our definition of scientific impact [36, 37], but also deepen our understanding of the role of science in human society. Taken together, these results unveil core patterns in the global policy landscape and its dynamics in the COVID-19 pandemic, opening new windows in our quantitative understanding and predictions of policy responses and the enormously important outcomes they govern.

# References

1. Lu, X., L. Bengtsson, and P. Holme, *Predictability of population displacement after the 2010 Haiti earthquake.* Proceedings of the National Academy of Sciences, U.S.A., 2012. **109**(29): p. 11576-11581.

2. Kryvasheyeu, Y., et al., *Rapid assessment of disaster damage using social media activity.* Science Advances, 2016. **2**(3): p. e1500779.

3. Bagrow, J.P., D. Wang, and A.-L. Barabasi, *Collective response of human populations to large-scale emergencies.* PLoS ONE, 2011. **6**(3): p. e17680.

4. Lazer, D., et al., *The parable of Google Flu: Traps in big data analysis.* Science, 2014. **343**(6176): p. 1203-1205.

5. Rourke, M., et al., *Policy opportunities to enhance sharing for pandemic research.* Science, 2020. **368**(6492): p. 716-718.

6. Hale, T., et al., *Variation in government responses to COVID-19*, in *Blavatnik School of Government Working Paper*. 2020.

7. Kraemer, M.U.G., et al., *The effect of human mobility and control measures on the COVID-19 epidemic in China.* Science, 2020. **368**(6490): p. 493-497.

8. Chinazzi, M., et al., *The effect of travel restrictions on the spread of the 2019 novel coronavirus (COVID-19) outbreak.* Science, 2020. **368**(6489): p. 395-400.

9. Anderson, R.M., et al., *How will country-based mitigation measures influence the course of the COVID-19 epidemic?* Lancet, 2020. **395**(10228): p. 931-934.

10. Flaxman, S., et al., *Estimating the effects of non-pharmaceutical interventions on COVID-19 in Europe.* Nature, 2020.

11. Hsiang, S., et al., *The effect of large-scale anti-contagion policies on the COVID-19 pandemic.* Nature, 2020.

12. Holtz, D., et al., *Interdependence and the cost of uncoordinated responses to COVID-19.* 2020.

13. Jia, J.S., et al., *Population flow drives spatio-temporal distribution of COVID-19 in China.* Nature, 2020.

14. Colizza, V., et al., *Modeling the worldwide spread of pandemic influenza: Baseline case and containment interventions.* PLoS Medicine, 2007. **4**(1): p. e13.

15. Johnson, N., et al., *Pattern in escalations in insurgent and terrorist activity.* Science, 2011. **333**(6038): p. 81-84.

16. Yin, Y., et al., *Quantifying the dynamics of failure across science, startups and security.* Nature, 2019. **575**(7781): p. 190-194.

17. Barabási, A.-L., *Bursts: The Hidden Patterns Behind Everything We Do, from Your E-mail to Bloody Crusades.* 2010, New York, NY, USA: Penguin.

18. Taleb, N.N., *The Black Swan: The Impact of the Highly Improbable.* Vol. 2. 2007: Random House.

19.  Azoulay, P. and B. Jones, *Beat COVID-19 through innovation.* Science, 2020. **368**(6491): p. 553.

20.  Van Bavel, J.J., et al., *Using social and behavioural science to support COVID-19 pandemic response.* Nature Human Behaviour, 2020. **4**: p. 460-471.

21.  Lazer, D., et al., *Computational social science.* Science, 2009. **323**(5915): p. 721-723.

22.  Editorial, *COVID-19 has reinforced the importance of preprints as an indispensable means for rapid research dissemination.* Nature Biotechnology, 2020. **38**: p. 507.

23.  Dong, E., H. Du, and L. Gardner, *An interactive web-based dashboard to track COVID-19 in real time.* Lancet Infectious Diseases, 2020. **20**: p. 533-534.

24.  Huang, C., et al., *Clinical features of patients infected with 2019 novel coronavirus in Wuhan, China.* Lancet, 2020. **395**(10223): p. 497-506.

25.  Ledford, H. and R. Van Noorden, *High-profile coronavirus retractions raise concerns about data oversight.* Nature, 2020. **582**(7810): p. 160.

26.  London, A.J. and J. Kimmelman, *Against pandemic research exceptionalism.* Science, 2020. **368**(6490): p. 476-477.

27.  Kitsak, M., et al., *Identification of influential spreaders in complex networks.* Nature Physics, 2010. **6**(11): p. 888-893.

28.  Albert, R., H. Jeong, and A.-L. Barabási, *Error and attack tolerance of complex networks.* Nature, 2000. **406**(6794): p. 378-382.

29.  Cohen, R., et al., *Resilience of the internet to random breakdowns.* Physical Review Letters, 2000. **85**(21): p. 4626.

30.  Holden , T.H., *Why WHO?* Science, 2020. **368**(6489): p. 341.

31.  Editorial, *Withholding funding from the World Health Organization is wrong and dangerous, and must be reversed.* Nature, 2020. **580**(7804): p. 431-432.

32.  Gibney, E., *Whose coronavirus strategy worked best? Scientists hunt most effective policies.* Nature, 2020. **581**(7806): p. 15-16.

33.  Financial Times. *Coronavirus tracked: Has your country's epidemic peaked?* . 2020 2020-05-30]; Available from: https://ig.ft.com/coronavirus-chart.

34.  Bento, A.I., et al., *Evidence from Internet search data shows information-seeking responses to news of local COVID-19 cases.* Proceedings of the National Academy of Sciences, U.S.A., 2020. **117**(21): p. 11220-11222.

35.  Kosack, S., et al., *Functional structures of US state governments.* Proceedings of the National Academy of Sciences, U.S.A., 2018. **115**(46): p. 11748-11753.

36.  Wang, D., C. Song, and A.-L. Barabási, *Quantifying long-term scientific impact.* Science, 2013. **342**(6154): p. 127-132.

37.  Fortunato, S., et al., *Science of science.* Science, 2018. **359**(6379): p. eaao0185.

**Acknowledgements** We thank Euan Adie and all members of the Center for Science of Science and Innovation (CSSI) at Northwestern University for their helpful discussions. This work uses data sourced from Overton.io and Dimensions.ai and is supported by the Air Force Office of Scientific Research under award number FA9550-17-1-0089 and FA9550-19-1-0354, National Science Foundation grant SBE 1829344, the Alfred P. Sloan Foundation G-2019-12485.

**Author contributions** D.W. conceived the project and designed the experiments; Y.Y. collected data; J.G. and Y.Y. performed empirical analyses with help from D.W. and B.F.J.; all authors discussed and interpreted results; all authors wrote and edited the manuscript.

**Competing interests** The authors declare no competing interests.

**Data availability** Deidentified data necessary to reproduce all plots and statistical analyses will be made freely available. COVID-19 cases and deaths data are publicly available. Those who wish to access raw data should contact the data sources directly.

**Code availability** Code will be made freely available.

**Figures captions**

**Figure 1. Policy responses to the COVID-19 pandemic. (A)** Linear growth in the cumulative number of all policy documents published since 2020. **(B)** Non-linear growth in the cumulative number of COVID-19 policy documents. Timeline shows major events including lockdowns and total confirmed cases of the COVID-19 infection. **(Inset)** Same as the main panel but in log-linear scale. Red dashed line shows an exponential function as a guide to the eye. **(C)** Policy documents mirror the case dynamics, showing a synchrony between the share of COVID-19 policy documents among all policy documents and the number of total confirmed cases. **(D)** The share of COVID-19 policy documents across three broad subject categories (21-day moving average). **(E)** The share of COVID-19 policy documents across topics (21-day moving average). Only the top 10 topics are shown (SI S2.1). Policy documents are counted multiple times by topics they cover. Color blue and red marks health- and economy-related topics, respectively. **(F)** Word clouds of all topics in COVID-19 policy documents published before **(F)** and after **(G)** the WHO pandemic declaration (March 11, 2020). Throughout the figures, the black dashed line marks the date of the WHO pandemic declaration.

**Figure 2. The use of science in COVID-19 policy documents. (A)** Probability of citing scientific references for COVID-19 policy documents published in 2020 (21-day moving average). The dashed red line represents the average rate of citing science by Non-COVID-19 policy documents in 2020. **(B)** Distribution of publication years of scientific papers cited by COVID-19 and other policy documents. There is an unusual spike in citing papers published in 2020, indicating that COVID-19 policies draw heavily on recent scientific evidence. **(C)** Research fields of all scientific papers cited by COVID-19 policy documents across three broad field categories (21-day moving average). There is a clear shift from the initial focus on health and science to economy and society, which is consistent with the results in Fig. 1D. **(D)** Among COVID-related scientific papers, those cited by COVID-19 policy documents on average have greater citation impact within science. **(E)** For different journals and preprint servers, the number of COVID-19 papers (x-axis) and the average number of citations from COVID-19 policy documents (y-axis) in 2020. While preprint servers published a large number of COVID-19 papers, peer-reviewed journal papers received substantially more COVID-19 policy citations than preprints. **(F)** COVID-19 policy documents that cite scientific papers are much more likely to be cited by other COVID-19 policy documents.

**Figure 3. The complex network behind COVID-19 policy documents. (A)** Number of COVID-19 policy documents published by governments, think tanks, and intergovernmental organizations (IGOs). **(B)** Probability of citing science for these different institutions. **(C)** Network visualization for the COVID-19 policy document citation network. Each node corresponds to a COVID-19 policy document, colored by the institution type to which it belongs, A link between documents is colored by a mixture between colors of the source and target nodes. The size of each node is proportional to the number of citations it receives from other COVID-19 policy documents. **(D)** The same as **(C)** but removing IGOs. **(E)** The same as **(C)** but only keeping policy documents published by government agencies (removing both IGOs and think tanks). In the absence of IGO policy documents, connectivity declines especially sharply and government policy documents become isolated in the citation network. **(F)** The citation distribution (in-degree) within the policy network by institutions. **(Inset)** The coreness score distribution ($k$-core). As the coreness measure is commonly used in undirected networks, here we treat the network as undirected when calculating the $k$-core score. **(G)** The weakly giant connected component size as we randomly remove policy documents from IGOs/think tanks/government agencies. Error bars represent standard errors. **(H)** Top 5 IGOs in the network, ranked by the total number of policy citations in the network. We also calculate their number of nodes (policy documents in the network).

**Figure 4. Policy attention and the effectiveness of local COVID-19 response. (A)** The significant negative correlation between the share of COVID-19 policy documents published in the 30 days before 3 daily deaths were first recorded (i.e., the prior 30 days), and the total deaths in the 30 days after 3 daily deaths were first recorded (i.e., the following 30 days). To accommodate the logarithmic scale, the share of COVID-19 policy documents is represented as ($\#COVID$+1)/($\#ALL$+1), where $\#COVID$ and $\#ALL$ are the number of COVID-19 and all policy documents, respectively. The gray line is a linear fit to the data. **(B)** The residual plot shows that a significant negative correlation remains between COVID-19 policy share in the prior 30 days (x-axis) and the residual of the regression (y-axis) after controlling for GDP per capita, population size, the calendar day of 3 daily deaths were first recorded, and a dummy variable capturing whether a country has no COVID-19 policy documents in the prior 30 days (SI S2.4). **(C)** The predictive power of the share of COVID-19 policy attention together with all other control variables in **(B)** (x-axis, the composite index) for total deaths in the following 30 days (y-axis).

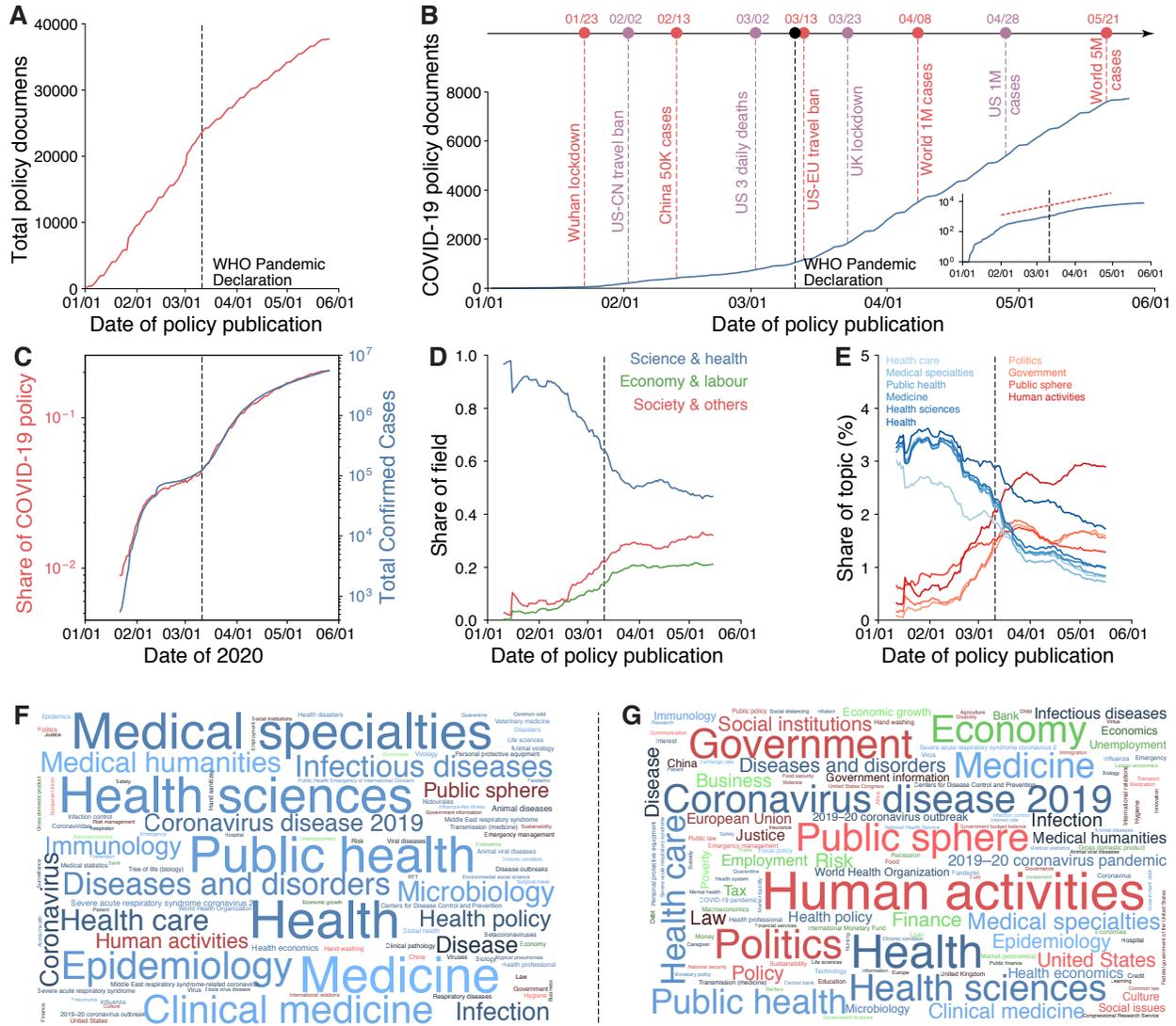

**Figure 1. Policy responses to the COVID-19 pandemic.**

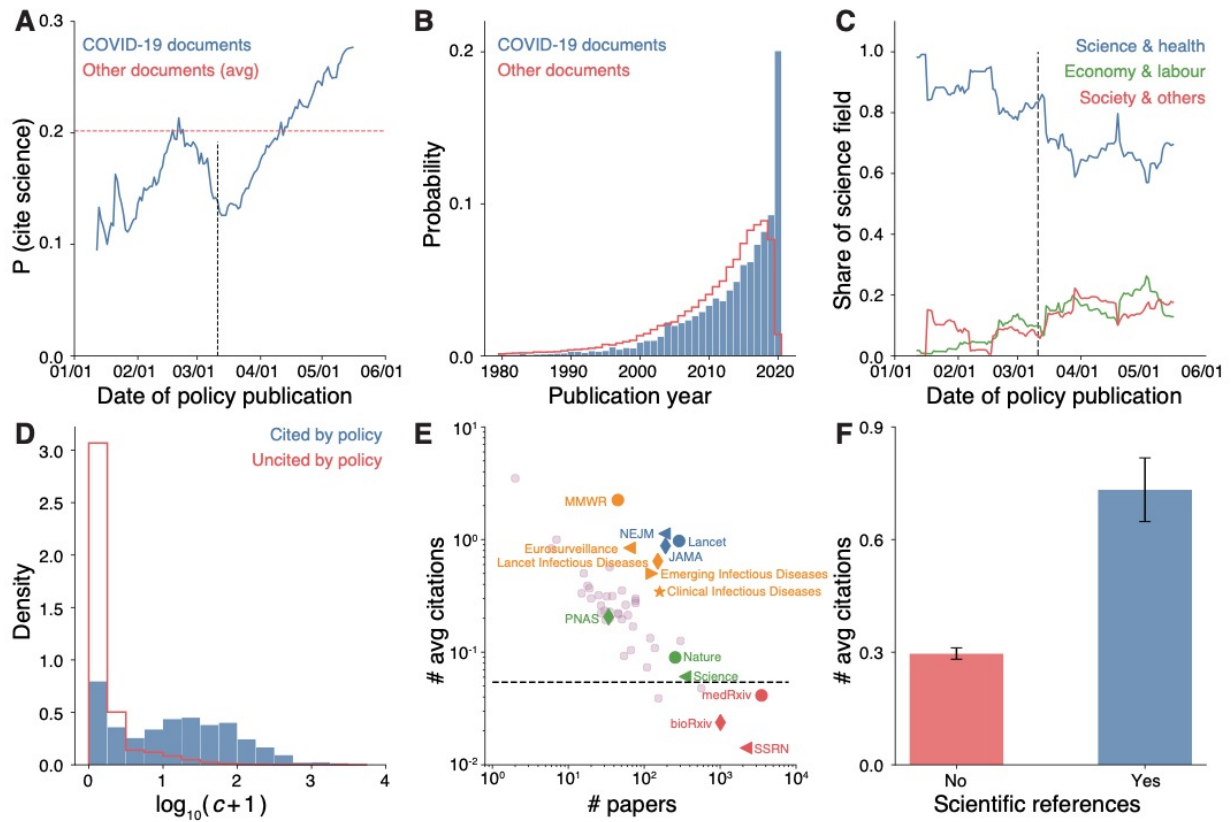

**Figure 2. The use of science in COVID-19 policy documents.**

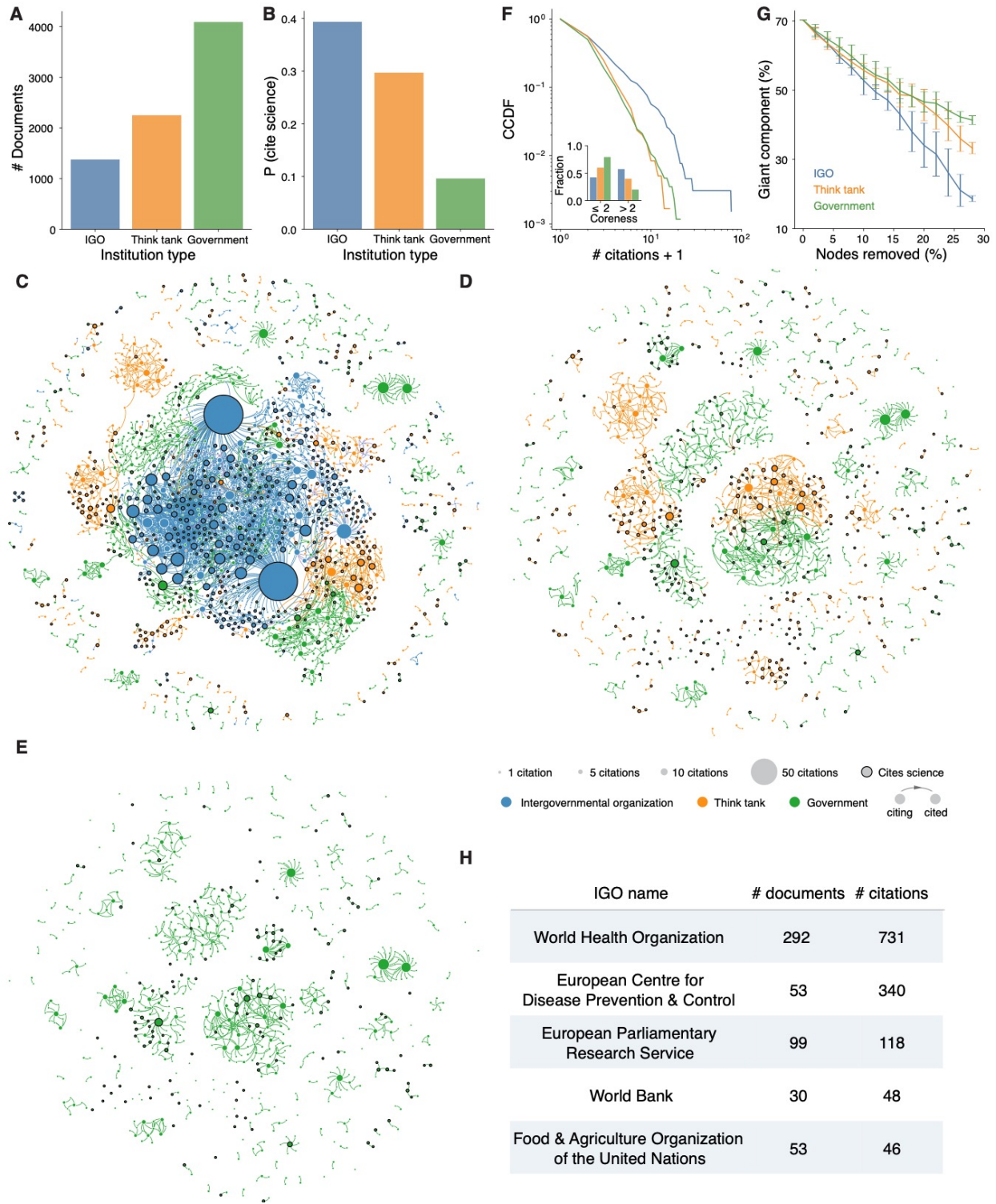

**Figure 3. The complex network behind COVID-19 policy documents.**

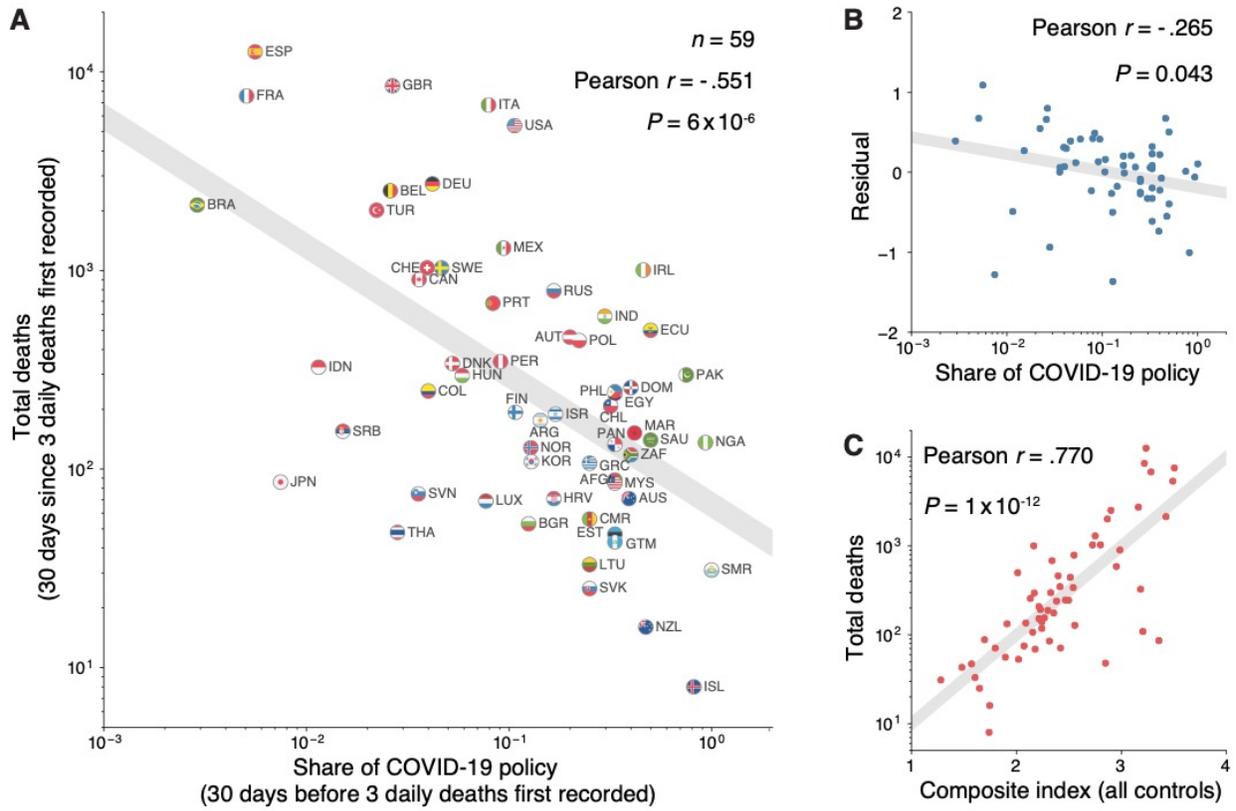

**Figure 4. Policy attention and the effectiveness of local COVID-19 response.**

# Supplementary Information for
# Quantifying Policy Responses to a Global Emergency: Insights from the COVID-19 Pandemic

Jian Gao[¶], Yian Yin[¶], Benjamin F. Jones[*], Dashun Wang[*]

[¶]These authors contributed equally to this work.

[*]Correspondence to: bjones@kellogg.northwestern.edu, dashun.wang@northwestern.edu.

**Table of Contents**

# S1 Data Description

## S1.1 Overton data overview

To understand the global policy response to the COVID-19 pandemic, we leverage a novel dataset provided by Overton (www.overton.io). Overton data captures among the world's largest collection of policy documents [1]. Policy documents are broadly defined as documents written primarily for or by policymakers, and include documents from government agencies, think tanks as well as intergovernmental organizations (IGOs) [1]. The data is updated weekly, allowing us to trace how policy responses evolve in nearly real time. In 2020, the Overton database has captured over 43,000 policy documents from 114 countries, collecting documents from more than 750 different sources worldwide.

In this paper, we use the full set of policy documents published from January 1st 2020 to May 26th 2020. We use an API to obtain policy documents from each policy source separately. For each document, we have information on its title, original URL, publication date, document type, policy source, and subject classification codes.

To identify COVID-19 related policy documents, we leverage Overton's technical capabilities which combine translation of policy document into English, with keyword-based search for a list of COVID-19 related keywords across multiple languages [2]. Note that the list of keywords also relies on pre-translation to English by Overton (see S1.4 below). While the data cover a large number of countries, there are two notable exceptions in our analyses. First is the Netherlands. One empirical limitation is that our access to the API only allows for at most 2,000 results per query, and there is one country that has exceeded this limit. Indeed, the Government of the Netherlands (Rijksoverheid) has published over 4,000 documents in 2020, making it impossible to collect the full set of its policy documents. We therefore exclude Netherlands from our analysis to avoid potential bias. Second is mainland China. Although COVID-19 originated from Wuhan, China, policy documents from mainland China are missing in the Overton database for 2020. Personal correspondence with the Overton team has not pinpointed the reason for this missing data, but it may have to do with web crawling issues on Chinese government websites [3].

We focus on policy documents published by (1) government agencies and think tanks in country members of the United Nations, and (2) intergovernmental organizations (defined by "IGO" and "EU" source labels in the Overton data). For the temporal coverage, we froze our data on May 26th, 2020. We noticed that there is an unusual number of documents published on January 1st 2020 (1,036 documents, about 4 times of a normal day), possibly due to default classification to this date. We further excluded these documents from our analyses to avoid including policy documents that have inaccurate publication dates. Lastly, since our main focus is on policy documents, we follow Overton's suggestion [3] and further filter on the document type by using only "publications" (95% of the total documents), removing other types such as "working papers", "transcripts", "blog posts", and "clinical guides".

In total, we analyzed 7,730 COVID-19 documents out of 37,725 documents published across 114 countries (including 402 think tanks), 55 IGOs (including 4 in EU).

## S1.2 Dimensions publication data

We further link the scientific references cited within the policy documents to a database of scientific publications. For this database, we use Dimensions [4], a recent data product by Digital Science. Each scientific reference from Overton has a unique DOI (Digital Object Identifier), one of the most commonly used identifiers for scientific publications. We retrieve papers from Dimensions API using the DOI information. We find that the vast majority of the references can be matched to Dimensions records (79,669 out of 84,964 papers, or 93.8% of the scientific references). For each paper we obtain information on its title, author list, affiliation(s), publishing venue, publication date, fields of study, references and citations received. Among the papers we analyzed, 79 of the 79,669 are matched to more than one record in Dimensions; hence to avoid duplications we keep the single item with most complete records, as determined by the number of references and citations.

We also constructed another set of COVID-19 related scientific publications by searching for papers published in 2020 with the following query suggested by Dimensions [5]:

> *"2019-nCoV" OR "COVID-19" OR "SARS-CoV-2" OR "HCoV-2019" OR "hcov" OR "NCOVID-19" OR "severe acute respiratory syndrome coronavirus 2" OR "severe acute respiratory syndrome corona virus 2" OR (("coronavirus" OR "corona virus") AND (Wuhan OR China OR novel)),* yielding in total 40,733 papers published in 2020 out of all articles indexed by Dimensions.

## S1.3 COVID-19 case and death tracking data

We use country-level daily statistics for COVID-19 confirmed cases and deaths from two independent sources. Our primary data source is the COVID-19 Data Repository provided by the Center for Systems Science and Engineering (CSSE) at Johns Hopkins University [6], which has been extensively used in COVID-19 related studies. To ensure the robustness of our results, we also repeated our analyses using an alternative dataset provided by the European Centre for Disease Prevention and Control (ECDC) [7], which includes not only information on COVID-19 confirmed cases and deaths, but also country-level covariates such as GDP per capita and population (see also S2.3). In our robustness checks (S3.4), we repeated our analyses using the COVID-19 confirmed deaths reported by ECDC, confirming that our results are not sensitive to the specific COVID-19 tracking data source.

## S1.4 Data limitations

Although our datasets represent among the largest collection of records in their respective domains, there are potential limitations of the data readers should keep in mind.

The Overton database assembles documents from a large number of sources, hence there might be potential sampling biases across different document types, countries and languages. For example, some countries may have more policy documents recorded in the data simply because their policy sources are more easily accessible, and documents better organized. Given the novelty of this database, we are not aware of other peer datasets that may help us assess the extent of potential sampling biases. Therefore, in our analyses, key results are based on comparisons between COVID-19 documents and all other documents recorded for that country or organization, i.e. the share of policy attention to COVID-19, which normalizes the total number of policy documents published by an entity.

Furthermore, COVID-19 related policy documents are identified through keyword searches, which may introduce both false positives and false negatives. Indeed, there have been several other notable coronavirus outbreaks in the past, such as SARS in 2003, indicating that our keyword search will also yield policy documents about other coronaviruses than COVID-19 (false positives). Yet, since our analysis only focuses on policy documents published in 2020, and all other coronavirus outbreaks took place before 2020, it suggests that the policy attention on coronavirus captured by our 2020 sample primarily concerns COVID-19. Second, false negatives (i.e., COVID-19 policy documents that are not captured by the keywords listed above) may also emerge from the keyword-based search approaches. Inspecting the official approach used by Overton for producing their own COVID-19 analysis [8] reveals that the keyword list contains several COVID-19 related words not only in English but also some Non-Latin languages (translations based on Wikidata [9]). Although this list is still not exhaustive, such efforts may help to reduce potential language biases.

Taken together, there are several potential sampling biases by countries and languages that readers should keep in mind. However, in the context of regression relationships (e.g., as in Fig. 4 of the main text), it is worth noting that noise in explanatory variable (e.g., the COVID-19 policy share) will attenuate the results toward zero, making it harder to find a significant statistical relationship. Indeed, as measurement errors are likely to introduce noise to our data, they suggest conservative biases to the regression relationships we observed. Related, in several different robustness checks, we find that if we focus on the portion of the data that has a higher reliability, the relationships we observe tend to strengthen further. For example, S3.7 presents robustness checks by separating English and non-English speaking countries, showing that if we only focus on English speaking countries (less bias by language), the correlations we observe become stronger. Similar trends are found in S3.8, where we limit the minimum number of COVID-19 policy documents published by each country, which again increases the overall correlation.

# S2 Methods

## S2.1 Field and topic classifications

We leverage field and topic classifications from Overton and Dimensions to determine the primary focus of each policy document and scientific paper.

Overton uses machine learning approaches to assign fields and topics to policy documents. The Overton policy field classification is primarily based on the International Press Telecommunications Council (IPTC) Subject Codes taxonomy, the global standards body of the news media. In our analysis, we use 18 top-level fields in Overton and further group them into three major field categories:

**Science & Health**: "health", "science and technology".
**Economy & Labour**: "economy, business and finance", "labour", "prices".
**Society & Others**: "arts, culture and entertainment", "conflicts, war and peace", "crime, law and justice", "disaster, accident and emergency incident", "education", "environment", "human interest", "lifestyle and leisure", "politics", "religion and belief", "society", "sport", "weather".

The Overton policy topic also implements a more fine-grained classification system. For example, in Fig. 1e, we used this classification scheme and show the top 10 topics by volume, after excluding one generic topic ("coronavirus disease 2019").

Dimensions implements the Fields of Research (FOR) classification for scientific papers. The FOR is a component of the Australian and New Zealand Standard Research Classification (ANZSRC) system, which follows a three-level hierarchy (divisions, groups and fields) and covers a broad set of research fields from the sciences and engineering, social sciences, and arts and humanities. In our analysis, we use 22 top-level divisions as the research fields of scientific papers. Analogously, as we do for policy documents, we further group 22 top-level divisions into the same three major field categories:

**Science & Health**: "Agricultural and Veterinary Sciences", "Biological Sciences", "Chemical Sciences", "Earth Sciences", "Engineering", "Environmental Sciences", "Information and Computing Sciences", "Mathematical Sciences", "Medical and Health Sciences", "Physical Sciences", "Psychology and Cognitive Sciences", "Technology".

**Economy & Labour**: "Commerce, Management, Tourism and Services", "Economics".

**Society & Others**: "Built Environment and Design", "Education", "History and Archaeology", "Language, Communication and Culture", "Law and Legal Studies", "Philosophy and Religious Studies", "Studies in Creative Arts and Writing", "Studies in Human Society".

## S2.2 Policy citation network

We construct a citation network using all COVID-19 related policy documents, where each directed link A→B corresponds to a citation from document A to document B. Self-citations (self-loops) are removed and only nodes with non-zero degree (i.e. documents that either cite or are cited by at least one other COVID-19 policy document) are visualized in the network (Fig. 3 of the main text).

To compare the importance of documents from various institutions, we simulate a random node removal procedure and study the relative size of the giant weakly connected component when a small fraction $f$ of the nodes is removed [10, 11]. For each institution type (IGO/think tank/government), we perform 50 independent simulations.

The relative importance of individual nodes is measured by $k$-core decomposition [12, 13], where $k$-core is defined as the maximal subgraph with all nodes having degree no smaller than $k$. A node has coreness of $k$ if it belongs to $k$-core but not $(k+1)$-core. Nodes with higher/lower coreness are located in the core/periphery of the network. We treat the network as undirected (defining degree as the sum of in-degree and out-degree) when calculating the coreness index.

## S2.3 Country-level policy response

We quantify the COVID-19 policy attention at the individual country level and explore its correlation with the country's effectiveness in containing the pandemic. For each country in our analysis, we mark the calendar date on which three deaths from COVID-19 in one day were first recorded [14]. We then calculate the share of COVID-19 policy documents among all policy documents in a country during the prior 30 days (i.e., in the 30 days before three daily deaths were first recorded in that country), which approximates the country's COVID-19 policy attention. We further measure the country's total deaths from COVID-19 over the following 30 days (i.e., in the 30 days after three daily deaths were first recorded in that country), which approximates the country's effectiveness in containing the pandemic. Both calculations are only possible if a country (1) has three daily deaths between January 31 and April 26, 2020; and (2) has at least one policy document published in the prior 30 days. After this filtering, there are 59 countries left, which we use as the full sample for Fig. 4 of the main text and for the various regression analyses (S2.4). This subset accounts for approximately 92.3% of confirmed deaths worldwide as of May 30th.

In addition to the share of COVID-19 policy documents, we consider an orthogonal COVID-19 policy attention measure, the stringency index, provided by the Oxford COVID-19 Government Response Tracker (OxCGRT) [15, 16]. Independent of the Overton policy document database that we rely on, OxCGRT collects information on 17 indicators of government responses to the COVID-19 pandemic such as travel restrictions and school closures. All these indicators are aggregated into an overall government response index (the stringency index), which records how the response of governments has varied over all indicators. The value of the stringency index is from 0 to 100 (from low to high level of government action). As of May 31, 2020, OxCGRT provides the daily stringency index for more than 160 countries. Consistent with our analysis in Fig. 4 of the main text, for each country we calculate an average stringency index during the prior 30 days (i.e., in the 30 days before three daily deaths were first recorded in that country), which serves as a separate proxy for a country's COVID-19 policy attention. The average stringency index is available for all the 59 countries that we analyze. We find that, although the stringency index measures policies that are specifically designed toward containing the pandemic, it exhibits a comparable level of predictive power as our measure (S3.10), which as an orthogonal source of policy data further confirms the robustness of our findings.

## S2.4 Regression methods

We employ an ordinary least squares (OLS) regression model to estimate the relations between the country-level COVID-19 policy attention and the country's effectiveness in containing the pandemic. The OLS regression takes the following form:

$$y_i = \beta_0 + \beta_1 x_{i,1} + \beta_2 x_{i,2} + \beta_3 x_{i,3} + \beta_4 x_{i,4} + \beta_5 x_{i,5} + \epsilon_i, \qquad \text{(Eq. 1)}$$

where $i$ indexes a country and $\epsilon_i$ is the error term. Other variables are defined as follows:

*Dependent variable*: The dependent variable $y_i = \log_{10} Death_i$ is the total COVID-19 confirmed deaths in the logarithmic scale in country $i$ in the 30 days after three daily deaths were first recorded in that country (i.e., the following 30 days), which approximates the country's effectiveness in containing the pandemic.

*Independent variable of interest*: The variable $x_{i,1}$ is the share of COVID-19 policy documents among all policy documents published by country $i$ in the 30 days before three daily deaths were first recorded (i.e., the prior 30 days) in that country. In our analysis, we use three different variations for the independent variable $x_{i,1}$:

1) $x_{i,1} = \#COVID_i / \#ALL_i$ is the ratio between the number of COVID-19 policy documents ($\#COVID_i$) and the number of all policy documents ($\#ALL_i$) published by country $i$ in the prior 30 days. Note that, to include country $i$ in the regression model, the country must have published at least one policy documents ($\#ALL_i >= 0$) in the prior 30 days and cover the entire following 30 days (see S2.3 for details).

2) $x_{i,1} = \log_{10}(\#COVID_i / \#ALL_i)$ is the share of COVID-19 policy documents, now measured on the logarithmic scale. Note that the logarithmic value is undefined when $\#COVID_i = 0$. The regression thus drops countries where the logarithmic value is undefined and only includes countries that published at least one COVID-19 policy document (i.e., $\#COVID_i >= 1$) in the prior 30 days. Alternatively, we consider this logarithmic regression model but with a dummy variable for country cases where $\#COVID_i = 0$. In this case, we continue to include all countries by resetting the logarithmic COVID share value to 0 and using the dummy variable to account for these otherwise dropped cases (see S3.4 for details).

3) $x_{i,1} = \log_{10}[(\#COVID_i + 1)/(\#ALL_i + 1)]$ is the COVID share in the logarithmic scale. As in 1), we add one to both the number of COVID-19 policy documents and the number of all policy documents before taking the log. This provides a simple alternative to avoid dropping cases where $\#COVID_i = 0$ and is visualized in Fig. 4 of the main text.

*Dummy variable*: The dummy variable $x_{i,2} = D_i^{COVID}$ indicates whether country $i$ has published COVID-19 policy documents in the prior 30 days. The value $D_i^{COVID} = 1$ if the number of COVID-19 policy documents published by country $i$ is 0 (i.e., $\#COVID_i = 0$) in the prior 30 days, and $D_i^{COVID} = 0$ if otherwise (i.e., $\#COVID_i >= 1$). When including the COVID share $x_{i,1} = \log_{10}(\#COVID_i/\#ALL_i)$ together with the dummy variable $x_{i,2} = D_i^{COVID}$ in the regression model, the dummy variable $x_{i,2}$ accounts for the effects of publishing no COVID-19 policy documents in the prior 30 days on the total deaths in the following 30 days.

*Other control variables*: In the regression model, we additionally control for three variables that may also have effects on a country's measured response. (1) $x_{i,3} = \log_{10}(GDPpc_i)$ is the GDP per capita of country $i$ in the logarithmic scale. (2) $x_{i,4} = \log_{10}(Pop_i)$ is the total population of country $i$ in the logarithmic scale. (3) $x_{i,5} = Day_i$ is the calendar day of 2020 on which three confirmed deaths from COVID-19 in one day were first recorded in country $i$.

# S3 Robustness checks

## S3.1 Temporal correlation between COVID-19 policy attention and the evolution of the pandemic

We compare the share of COVID-19 policy attention and total confirmed deaths from COVID-19 over time, finding a high degree of similarity between the two trajectories (Fig. S2A). To test if there are systematic delays between the share of COVID-19 policy attention and the pandemic progress (both cases and deaths), we further calculate the correlation between the two time-series. To account for the approximate exponential growth of both curves, we take the first-order difference of logarithm transformation, defined as

$$Y_t = \log_{10} X_{t+1} - \log_{10} X_t.$$

Correlations between the transformed series suggest that the relationship between the time series is closest when the time offset, $\Delta t$, between the share of COVID-19 policy documents and COVID-19 deaths is close to 0 (see Fig. S2B for deaths and Fig. S2C for cases), suggesting a high degree of synchronicity between the share of COVID-19 policy attention and how the pandemic evolves.

### S3.2 Evolution of fields and topics of policy documents and cited papers

In Fig. 1D of the main text, we group COVID-19 related policy documents into three major field categories (see also S2.1) and observe clear shifts in policy attention related to the pandemic, with policy priorities shifting from issues in public health to economy and society. At the narrower topic level, in Fig. 1e and 1f we find a clear decrease in the share of health-related topics and an increase of topics related to economy and society since early March 2020.

Here, we test the robustness of these observations by looking into data by month and for individual fields. We find similar shifts in topics of COVID-19 policy documents around March 2020 (Fig. S3A), with January and February primarily focusing on public health and medicine while April and May showing more expansive subject orientations and increased focus on social and economic issues. Further, we calculate the share of COVID-19 policy documents by each field and plot the results for the top 10 fields ranked by total COVID-19 policy documents (Fig. S3B), finding similar shifts in COVID-19 policy attention from health to economy and society. Similarly, we further consider the share of scientific papers cited by COVID-19 policy documents for individual Fields of Research (FOR). We find a clear shift of COVID-19 policy documents from drawing primarily upon biomedical literature to citing economics, society, and other FOR (see Fig. S3C for the top 10 FOR), showing the robustness of our results reported in Fig. 2C of the main text.

Furthermore, we repeat our analyses on the evolution of fields for other (i.e., non-COVID-19) policy documents published in the same period. We find that the shares of policy documents by three major field categories (Fig. S4A) and individual fields (Fig. S4B) stay relatively stable over time, suggesting that the shifts are observed in COVID-19 policy documents only.

## S3.3 Robustness checks on the use of science and policy document impact

Figure 2F of the main text documents a positive relationship between the use of scientific references and the number of citations from other COVID-19 policy documents, with policies that reference science receiving 0.436 additional policy citations on average (more than doubling the baseline rate). Here we further test the robustness of this result in a linear regression model to control for the publication date of the document as well as fixed effects for the country and type of institution, finding that all else being equal, COVID-19 policy documents that cite scientific papers are associated with 0.322 more policy citations than those that do not ($P < 0.001$). We also separate citations they receive into citations from the same institution or different policy institutions, finding additional citations for both types (0.184 more policy citations from the same institution and 0.137 more policy citations from the other institutions).

**Table S1.** Regressions considering the effects of policy publication date, the country, and the type of the institution.

| Variables | Dependent variable: Policy citations | | | | |
|---|---|---|---|---|---|
| | total | | | same institution | different institution |
| | (1) | (2) | (3) | (4) | (5) |
| D_cites_science | 0.436*** | 0.516*** | 0.322*** | 0.184*** | 0.137*** |
| | (0.053) | (0.052) | (0.054) | (0.040) | (0.023) |
| Policy date | | -0.010*** | -0.011*** | -0.007*** | -0.004*** |
| | | (0.001) | (0.001) | (0.001) | (0.000) |
| Institution type | | | Yes | Yes | Yes |
| Institution country | | | Yes | Yes | Yes |
| Constant | 0.297*** | 1.302*** | 1.097* | 0.732* | 0.365 |
| | (0.240) | (0.078) | 0.585 | (0.431) | (0.250) |
| Observations | 7730 | 7730 | 7730 | 7730 | 7730 |
| Adj. R2 | 0.009 | 0.031 | 0.065 | 0.045 | 0.055 |
| RMSE | 1.877 | 1.855 | 1.823 | 1.344 | 0.779 |
| *F* statistic | 68.84 | 125.61 | 9.48 | 6.83 | 8.09 |

Notes: Column (4) and (5) shows the results on using citations from same/different sources only. Standard errors in parentheses. Significant level: *p<0.1; **p<0.05; ***p<0.01.

## S3.4 Robustness checks on the relations between the share of policy attention and a country's effectiveness in response

We explore the relationship between a country's COVID-19 policy response and its effectiveness in containing the pandemic by employing the ordinary least squares (OLS) regression model given in Eq. (1), as described in S2.4. The dependent variable $\log_{10}(Death)$ is the logarithmic count of the country's total deaths from COVID-19 in the 30 days after 3 daily deaths were first recorded in that country. The independent variable of interest is the share of COVID-19 policy attention given by three alternative functional forms, namely, the linear share of COVID-19 policy documents, $\#COVID/\#ALL$; the logarithmic share of COVID policy documents, $\log_{10}(\frac{\#COVID}{\#ALL})$; and the logarithmic COVID share after adding 1 for both numbers of COVID and all policy documents: $\log_{10}(\frac{\#COVID+1}{\#ALL+1})$ (see S2.4 for more detail). Here, together with each of these functional forms for the independent variable of interest, we gradually include other control variables in the regression.

**Table S2.** Regressions considering the effects of COVID-19 policy attention on total deaths.

| Variables | Dependent variable: *Death* (Log10) | | | | | | | | |
|---|---|---|---|---|---|---|---|---|---|
| | (1) | (2) | (3) | (4) | (5) | (6) | (7) | (8) | (9) |
| $\#COVID/\#ALL$ | -0.844** | -1.615*** | -0.637* | | | | | | |
| | (0.388) | (0.403) | (0.368) | | | | | | |
| $\#COVID/\#ALL$ (Log10) | | | | -0.669*** | -0.669*** | -0.323** | | | |
| | | | | (0.170) | (0.164) | (0.140) | | | |
| $(\#COVID+1)/(\#ALL+1)$ (Log10) | | | | | | | -0.651*** | -0.662*** | -0.319** |
| | | | | | | | (0.130) | (0.125) | (0.127) |
| $GDPpc$ (Log10) | | | 0.372 | | | 0.412 | | | 0.340 |
| | | | (0.272) | | | (0.256) | | | (0.260) |
| $Population$ (Log10) | | | 0.427*** | | | 0.441*** | | | 0.411*** |
| | | | (0.111) | | | (0.104) | | | (0.106) |
| $Day$ | | | -0.021*** | | | -0.019*** | | | -0.017** |
| | | | (0.007) | | | (0.007) | | | (0.007) |
| $D^{COVID}$ | | -0.713*** | -0.263 | | 0.332 | 0.206 | | -0.368** | -0.163 |
| | | (0.187) | (0.178) | | (0.230) | (0.181) | | (0.150) | (0.139) |
| Constant | 2.530*** | 2.953*** | -0.424 | 1.908*** | 1.908*** | -1.316 | 1.821*** | 1.979*** | -0.916 |
| | (0.102) | (0.144) | (2.163) | (0.205) | (0.197) | (1.997) | (0.144) | (0.152) | (1.998) |
| Observations | 59 | 59 | 59 | 32 | 59 | 59 | 59 | 59 | 59 |
| Adj. R2 | 0.060 | 0.241 | 0.528 | 0.318 | 0.248 | 0.546 | 0.292 | 0.349 | 0.554 |
| RMSE | 0.688 | 0.618 | 0.488 | 0.640 | 0.615 | 0.478 | 0.597 | 0.573 | 0.474 |
| *F* Statistic | 4.732 | 10.23 | 13.97 | 15.46 | 10.56 | 14.97 | 24.89 | 16.56 | 15.41 |

Notes: The dummy variable $D^{COVID}= 1$ if $\#COVID = 0$, and $D^{COVID}= 0$ if otherwise. Countries where $\#COVID = 0$ are dropped in column (4), and these countries are included in column (5) after resetting the value of $\log_{10}(\#COVID/\#ALL)$ to 0. Standard errors in parentheses. Significant level: *p<0.1; **p<0.05; ***p<0.01.

In columns (1-3) of Table S2, we use the COVID share $\#COVID/\#ALL$ as a proxy for the share of COVID-19 policy attention. We find that this COVID-19 policy attention share by itself exhibits a significant, negative correlation with total deaths (see column (1)). Further, in column (2) we include the dummy variable $D^{COVID}$ that indicates whether a country has yet published COVID-19 policy documents and find the negative correlation further improves. In column (3) we include three other control variables, namely, GDP per capita, population, and the calendar day of 3 daily deaths were first recorded for each country. We find that the COVID share still has a significant predictive power for the total deaths after controlling for all other variables.

In columns (4-5), we use the COVID share in the logarithmic scale $\log_{10}(\frac{\#COVID}{\#ALL})$ as a proxy for the share of COVID policy attention. When including $\log_{10}(\frac{\#COVID}{\#ALL})$ solely in column (4), the regression drops countries with $\#COVID = 0$ (i.e., where $\log_{10}(\frac{\#COVID}{\#ALL})$ is undefined). In columns (5-6), we include all countries after resetting $\log_{10}\left(\frac{\#COVID}{\#ALL}\right) = 0$ for countries where $\#COVID = 0$ and use the dummy variable $D^{COVID}$ to account for publishing no COVID-19 policy documents. We find that the negative correlation between the COVID share and total deaths remain strong and significant.

In columns (7-9), we repeat the regression by using $\log_{10}(\frac{\#COVID+1}{\#ALL+1})$ as a proxy for the share of COVID policy attention. We find that the negative correlations remain robust, with a 10 percent increase in the adjusted COVID-19 policy share variable predicting 3.19 percent fewer deaths. GDP per capita has no significant effect on total deaths, whereas population size has a positive effect. Moreover, the calendar day of the first 3 daily deaths has significantly negative effects, suggesting that a later onset of COVID-19 outbreak is associated with fewer deaths in the following 30 days. These variables together can explain up to 55.4% of the variation in total deaths.

### S3.5 Thresholds on the minimum number of policy documents

We further check the robustness of the negative correlation between the COVID-19 policy attention in the prior 30 days and the total deaths in the following 30 days (i.e., the 30 days after 3 daily deaths were first recorded), by restricting the analysis to a minimum threshold number of policy documents that a country published in the prior 30 days (i.e., the 30 days before 3 daily deaths were first recorded),

Table S3 summarizes the results of regressions where we only include countries that published at least 10 policy documents in the prior 30 days. We find that the negative correlations remain largely significant. In particular, we notice from column (9) that the regression coefficient of the COVID share increases to -0.375 (from -0.319 in Table S2), suggesting that that COVID share exhibits a slightly stronger predictive power for total deaths in countries that published more policy documents.

**Table S3.** Regressions considering the effects of COVID-19 policy attention on total deaths for countries that published at least 10 policy documents in the prior 30 days.

| Variables | Dependent variable: $Death$ (Log10) | | | | | | | | #ALL >= 10 |
| | (1) | (2) | (3) | (4) | (5) | (6) | (7) | (8) | (9) |
|---|---|---|---|---|---|---|---|---|---|
| $\#COVID / \#ALL$ | -1.565*** | -1.866*** | -0.947 | | | | | | |
| | (0.517) | (0.553) | (0.566) | | | | | | |
| $\#COVID / \#ALL$ (Log10) | | | | -0.661*** | -0.661*** | -0.377** | | | |
| | | | | (0.200) | (0.198) | (0.182) | | | |
| $(\#COVID+1) / (\#ALL+1)$ (Log10) | | | | | | | -0.604*** | -0.746*** | -0.375* |
| | | | | | | | (0.181) | (0.194) | (0.203) |
| $GDPpc$ (Log10) | | | 0.718 | | | 0.890* | | | 0.805 |
| | | | (0.559) | | | (0.521) | | | (0.537) |
| $Population$ (Log10) | | | 0.567*** | | | 0.597*** | | | 0.552*** |
| | | | (0.196) | | | (0.188) | | | (0.196) |
| $Day$ | | | -0.011 | | | -0.010 | | | -0.010 |
| | | | (0.012) | | | (0.012) | | | (0.012) |
| $D^{COVID}$ | | -0.384 | -0.104 | | 0.699** | 0.516* | | -0.466* | -0.130 |
| | | (0.273) | (0.266) | | (0.334) | (0.290) | | (0.268) | (0.264) |
| Constant | 2.855*** | 3.000*** | -3.722 | 1.917*** | 1.917*** | -5.437 | 1.930*** | 1.889*** | -4.689 |
| | (0.134) | (0.168) | (4.297) | (0.257) | (0.255) | (4.065) | (0.239) | (0.233) | (4.143) |
| Observations | 37 | 37 | 37 | 27 | 37 | 37 | 37 | 37 | 37 |
| Adj. R2 | 0.185 | 0.207 | 0.427 | 0.277 | 0.202 | 0.451 | 0.221 | 0.263 | 0.437 |
| RMSE | 0.690 | 0.680 | 0.578 | 0.688 | 0.682 | 0.566 | 0.675 | 0.656 | 0.573 |
| $F$ Statistic | 9.167 | 5.703 | 6.364 | 10.95 | 5.569 | 6.911 | 11.18 | 7.423 | 6.592 |

Notes: The dummy variable $D^{COVID} = 1$ if $\#COVID = 0$, and $D^{COVID} = 0$ if otherwise. Countries where $\#COVID = 0$ are dropped in column (4), and these countries are included in column (5) after resetting the value of $\log_{10}(\#COVID/\#ALL)$ to 0. Standard errors in parentheses. Significant level: *p<0.1; **p<0.05; ***p<0.01.

In Table S4, we further limit our regression analysis to countries that published at least 25 policy documents in the prior 30 days. We notice that the regression coefficient of COVID share in column (9) further increases to -0.467, showing that the share of COVID-19 policy attention in the prior 30 days in countries that published more policy documents has again a steeper relationship with the total deaths in the following 30 days. With the smaller number of country observations that result from these policy document count restrictions, the standard errors tend naturally to increase in these regressions, yet most specifications remain statistically significant. Overall, these analyses tend to further support the robustness of our findings.

**Table S4.** Regressions considering the effects of COVID policy attention on total deaths for countries that published at least 25 policy documents in the prior 30 days.

| Variables | Dependent variable: *Death* (Log10) | | | | | | | | #ALL >= 25 |
|---|---|---|---|---|---|---|---|---|---|
| | (1) | (2) | (3) | (4) | (5) | (6) | (7) | (8) | (9) |
| #COVID / #ALL | -1.954* | -2.403** | -1.365 | | | | | | |
| | (1.01) | (1.082) | (1.167) | | | | | | |
| #COVID / #ALL (Log10) | | | | -0.558** | -0.558** | -0.369 | | | |
| | | | | (0.239) | (0.251) | (0.235) | | | |
| (#COVID +1) / (#ALL +1) (Log10) | | | | | | | -0.528** | -0.759*** | -0.467* |
| | | | | | | | (0.229) | (0.255) | (0.257) |
| GDPpc (Log10) | | | 0.956 | | | 0.990 | | | 0.888 |
| | | | (0.842) | | | (0.816) | | | (0.789) |
| Population (Log10) | | | 0.561* | | | 0.536* | | | 0.478* |
| | | | (0.297) | | | (0.282) | | | (0.274) |
| Day | | | -0.012 | | | -0.015 | | | -0.014 |
| | | | (0.019) | | | (0.016) | | | (0.016) |
| $D^{\text{COVID}}$ | | -0.410 | -0.179 | | 0.579 | 0.456 | | -0.666* | -0.353 |
| | | (0.366) | (0.341) | | (0.460) | (0.449) | | (0.371) | (0.352) |
| Constant | 2.934*** | 3.063*** | -4.648 | 2.074*** | 2.074*** | -5.063 | 2.058*** | 1.898*** | -4.330 |
| | (0.171) | (0.206) | (6.755) | (0.329) | (0.347) | (6.522) | (0.325) | (0.325) | (6.249) |
| Observations | 29 | 29 | 29 | 23 | 29 | 29 | 29 | 29 | 29 |
| Adj. R2 | 0.089 | 0.098 | 0.318 | 0.169 | 0.098 | 0.347 | 0.134 | 0.200 | 0.368 |
| RMSE | 0.746 | 0.742 | 0.646 | 0.706 | 0.742 | 0.632 | 0.727 | 0.699 | 0.621 |
| *F* Statistic | 3.742 | 2.516 | 3.610 | 5.463 | 2.519 | 3.976 | 5.327 | 4.495 | 4.263 |

Notes: The dummy variable $D^{COVID} = 1$ if $\#COVID = 0$, and $D^{COVID} = 0$ if otherwise. Countries where $\#COVID = 0$ are dropped in column (4), and these countries are included in column (5) after resetting the value of $\log_{10}(\#COVID / \#ALL)$ to 0. Standard errors in parentheses. Significant level: *p<0.1; **p<0.05; ***p<0.01.

## S3.6 Alternative measures for effectiveness in response and other data sources

In Fig. 4A of the main text, we presented the negative correlation between the share of COVID-19 policy attention in the prior 30 days and the total confirmed deaths in the following 30 days, where we use the COVID-19 death tracking data provided by Johns Hopkins University (JHU). Here, we consider alternative measures and show that our results are robust when using total confirmed cases (Fig. S5A; columns (1-2) of Table S5) and confirmed deaths per million population of a country (Fig. S5B; columns (3-4) of Table S5) in the following 30 days.

We further leverage a new dataset for daily COVID-19 case and death statistics, provided by the European Centre for Disease Prevention and Control (ECDC). Country-level death statistics collected by JHU and ECDC are highly correlated with each other. For example, the Pearson correlation coefficient between the total deaths (in the logarithmic scale) in the following 30 days recorded by the two data sources is 0.996. Here, we show that our results remain robust when using ECDC death statistics (Fig. S5D; columns (5-6) of Table S5).

**Table S5.** Regressions considering the effects of COVID policy attention on the effectiveness in containing the pandemic.

| Variables | Dependent variable: (Log10) | | | | | | | |
|---|---|---|---|---|---|---|---|---|
| | Cases (JHU) | | Deaths per Million (JHU) | | Deaths (ECDC) | | Deaths (JHU, English) | |
| | (1) | (2) | (3) | (4) | (5) | (6) | (7) | (8) |
| $(\#COVID+1)/(\#ALL+1)$ | -0.493*** | -0.096 | -0.301* | -0.319** | -0.555*** | -0.219* | -1.107** | -0.112 |
| (Log10) | (0.133) | (0.118) | (0.160) | (0.127) | (0.139) | (0.129) | (0.365) | (0.394) |
| $GDPpc$ (Log10) | | 0.541** | | 0.340 | | 0.292 | | -0.451 |
| | | (0.241) | | (0.260) | | (0.267) | | (0.560) |
| $Population$ (Log10) | | 0.520*** | | -0.589*** | | 0.403*** | | -0.100 |
| | | (0.098) | | (0.106) | | (0.110) | | (0.338) |
| $Day$ | | -0.017** | | -0.017** | | -0.022*** | | -0.061** |
| | | (0.006) | | (0.007) | | (0.008) | | (0.023) |
| $D^{COVID}$ | | -0.113 | | -0.163 | | -0.210 | | -0.983* |
| | | (0.129) | | (0.139) | | (0.139) | | (0.431) |
| Constant | 3.369*** | -1.031 | 0.830*** | 5.084** | 1.881*** | -0.122 | 1.792*** | 10.29 |
| | (0.146) | (1.853) | (0.176) | (1.998) | (0.156) | (2.131) | (0.271) | (6.613) |
| Observations | 59 | 59 | 59 | 59 | 59 | 59 | 14 | 14 |
| Adj. R2 | 0.181 | 0.572 | 0.042 | 0.597 | 0.204 | 0.536 | 0.387 | 0.731 |
| RMSE | 0.608 | 0.439 | 0.731 | 0.474 | 0.633 | 0.483 | 0.592 | 0.393 |
| F Statistic | 13.79 | 16.47 | 3.551 | 18.21 | 15.90 | 14.41 | 9.213 | 8.057 |

Notes: The dummy variable $D^{COVID} = 1$ if $\#COVID = 0$, and $D^{COVID} = 0$ if otherwise. Standard errors in parentheses. Significant level: *p<0.1; **p<0.05; ***p<0.01.

## S3.7 Results for English-speaking countries

To address possible concerns on the potential bias by language in Overton policy documents, we further performed robustness checks by separating English and non-English speaking countries in our analysis. Based on whether English is an official language [17], we consider a combination of three sets of countries as English-speaking countries in our analysis: (1) Countries where English is a de facto (i.e., practices that exist in reality) official language; (2) Countries where English is a de jure (i.e., practices that are legally recognized) and de facto official language; (3) Countries where English is a de facto official, but not primary language.

Out of all 59 countries in our analysis for Fig. 4A of the main text, 14 countries are English speaking countries. If we only focus on English speaking countries, the Pearson correlation between the share of COVID-19 policy attention in the prior 30 days and the total deaths in the following 30 days increases from -0.551 (Fig. 4A of the main text) to -0.659 (Fig. S5C; columns (7-8) of Table S5), showing that the correlation becomes stronger and remains significant ($P = 0.01$) without other controls, and despite the small sample size, suggesting the robustness of the results.

## S3.8 Thresholds on the minimal number of COVID-19 policy documents

We further focus on the portion of the data that has a higher reliability in our analysis by limiting countries that have published a minimum threshold number of COVID-19 policy documents in the prior 30 days. We find that the negative correlation between the share of COVID-19 policy attention and total deaths that we observe tends to strengthen. Specifically, when increasing the minimum number of COVID-19 policy documents from 1 to 10, the Pearson correlation coefficient increases from -0.63 to -0.78 (Fig. S6), and the correlation remains largely significant when controlling for other variables (Table S6).

**Table S6.** Regressions considering the effects of COVID policy attention on total deaths for countries that published a minimum number of COVID policy documents in the prior 30 days.

| Variables | Dependent variable: *Death* (Log10) | | | | | | | | | | | |
|---|---|---|---|---|---|---|---|---|---|---|---|---|
| | #COVID >= 1 | | #COVID >= 2 | | #COVID >= 3 | | #COVID >= 4 | | #COVID >= 5 | | #COVID >= 10 | |
| | (1) | (2) | (3) | (4) | (5) | (6) | (7) | (8) | (9) | (10) | (11) | (12) |
| (#COVID+1)/(#ALL+1) (Log10) | -0.783*** | -0.361* | -1.073*** | -0.745*** | -1.150*** | -0.758*** | -1.171*** | -0.556** | -1.300*** | -0.648* | -1.361*** | -0.218 |
| | (0.175) | (0.184) | (0.191) | (0.238) | (0.204) | (0.254) | (0.221) | (0.245) | (0.278) | (0.338) | (0.399) | (0.572) |
| *GDPpc* (Log10) | | 0.616 | | 0.225 | | 0.327 | | 0.295 | | 0.374 | | -1.117 |
| | | (0.452) | | (0.486) | | (0.536) | | (0.482) | | (0.649) | | (1.320) |
| *Population* (Log10) | | 0.509*** | | 0.399** | | 0.402** | | 0.445** | | 0.424** | | -0.375 |
| | | (0.155) | | (0.158) | | (0.166) | | (0.152) | | (0.183) | | (0.625) |
| *Day* | | -0.015 | | -0.009 | | -0.008 | | -0.022** | | -0.019 | | -0.075 |
| | | (0.010) | | (0.010) | | (0.010) | | (0.010) | | (0.014) | | (0.042) |
| Constant | 1.870*** | -3.052 | 1.738*** | -1.242 | 1.657*** | -1.794 | 1.664*** | -0.655 | 1.577*** | -1.163 | 1.737*** | 16.49 |
| | (0.192) | (3.313) | (0.194) | (3.362) | (0.208) | (3.637) | (0.221) | (3.318) | (0.247) | (4.429) | (0.329) | (13.77) |
| Observations | 32 | 32 | 25 | 25 | 23 | 23 | 20 | 20 | 16 | 16 | 10 | 10 |
| Adj. R2 | 0.379 | 0.584 | 0.559 | 0.683 | 0.583 | 0.679 | 0.588 | 0.770 | 0.582 | 0.725 | 0.542 | 0.713 |
| RMSE | 0.611 | 0.499 | 0.547 | 0.464 | 0.553 | 0.485 | 0.574 | 0.428 | 0.566 | 0.460 | 0.588 | 0.466 |
| *F* Statistic | 19.90 | 11.90 | 31.47 | 13.91 | 31.78 | 12.66 | 28.10 | 16.94 | 21.92 | 10.88 | 11.66 | 6.599 |

Notes: Standard errors in parentheses. Significant level: *p<0.1; **p<0.05; ***p<0.01.

## S3.9 Robustness checks on the onset of outbreaks

As a reference point approximating the onset of the outbreak, the paper takes the day where 3 daily deaths were first recorded. Here we present robustness checks on this reference point (Fig. S7; Table S7). Specifically, when taking the day where the first death was recorded in each country as the reference point [15, 16], a total of 82 countries are included in the analysis, and we find a significantly negative correlation (-0.550) between COVID share and total deaths (Fig. S7A; columns (1-2) of Table S7). Taking the reference point as the day when 5 daily deaths were first recorded (Fig. S7B; columns (3-4) of Table S7) and the day where 10 daily deaths were first recorded (Fig. S7C; columns (5-6) of Table S7), the negative correlation between COVID share and total deaths remains strong and significant, again confirming the robustness of our results.

**Table S7.** Regressions considering the effects of COVID policy attention on total deaths in the prior 30 days under different reference points.

| Variables | Dependent variable: *Death* (Log10) | | | | | | | |
|---|---|---|---|---|---|---|---|---|
| | First 1 Daily Death | | First 5 Daily Deaths | | First 10 Daily Deaths | | Total 100 Deaths | |
| | (1) | (2) | (3) | (4) | (5) | (6) | (7) | (8) |
| (#COVID+1)/(#ALL+1) | -0.852*** | -0.452*** | -0.680*** | -0.370*** | -0.570*** | -0.193 | -0.802*** | -0.285* |
| (Log10) | (0.144) | (0.141) | (0.134) | (0.133) | (0.154) | (0.136) | (0.160) | (0.145) |
| *GDPpc* (Log10) | | 0.734*** | | 0.308 | | 0.992*** | | 0.454 |
| | | (0.213) | | (0.292) | | (0.322) | | (0.284) |
| *Population* (Log10) | | 0.558*** | | 0.420*** | | 0.568*** | | 0.438*** |
| | | (0.105) | | (0.113) | | (0.141) | | (0.133) |
| *Day* | | -0.002 | | -0.013 | | -0.015** | | -0.031*** |
| | | (0.008) | | (0.008) | | (0.007) | | (0.008) |
| $D^{COVID}$ | | -0.218 | | -0.161 | | 0.195 | | 0.135 |
| | | (0.147) | | (0.143) | | (0.174) | | (0.165) |
| Constant | 1.049*** | -5.455*** | 1.984*** | -1.039 | 2.362*** | -4.617* | 2.331*** | 0.247 |
| | (0.150) | (1.808) | (0.142) | (2.267) | (0.174) | (2.519) | (0.156) | (2.354) |
| Observations | 82 | 82 | 53 | 53 | 40 | 40 | 39 | 39 |
| Adj. R2 | 0.294 | 0.526 | 0.322 | 0.543 | 0.246 | 0.587 | 0.390 | 0.694 |
| RMSE | 0.727 | 0.595 | 0.581 | 0.478 | 0.561 | 0.416 | 0.544 | 0.385 |
| F Statistic | 34.76 | 19.01 | 25.70 | 13.34 | 13.75 | 12.07 | 25.25 | 18.24 |

Notes: The dummy variable $D^{COVID} = 1$ if $\#COVID = 0$, and $D^{COVID} = 0$ if otherwise. *Day* is the calendar day of the reference point in 2020. Standard errors in parentheses. Significant level: *p<0.1; **p<0.05; ***p<0.01.

Besides the number of daily deaths, we use an alternative method by taking the day where a cumulative number of 100 deaths was first recorded as the reference point. After checking the relations between the share of COVID-19 policy documents in the 30 days before total 100 deaths was first recorded in a country and the total deaths after total 100 deaths was first recorded in that country, we find that the negative correlation further improves and remains significant (Fig. S7D; columns (7-8) of Table S7). Overall, these results support the robustness of our results.

### S3.10 Comparison with the Stringency Index

Finally, we leverage an independent measure on COVID-19 policy, the stringency index provided by the Oxford COVID-19 Government Response Tracker (OxCGRT) [15, 16], to further test the robustness of our finding on the relationship between COVID-19 policy attention and the effectiveness in containing the pandemic. For each country we calculate an average stringency index during the prior 30 days and employ it to predict the total deaths in the following 30 days. Table S8 summaries the results of regressions that include both the share of COVID-19 policy documents and the average stringency index together with other control variables. In order to compare the values of regression coefficients, here we take the z-score for all variables.

**Table S8.** Regressions considering the effects of both the share of COVID-19 policy documents and the average stringency index on total deaths.

| Variables | Dependent variable: $Death$ (Log10) | | | | | | | | | | | | | | |
|---|---|---|---|---|---|---|---|---|---|---|---|---|---|---|---|
| | (1) | (2) | (3) | (4) | (5) | (6) | (7) | (8) | (9) | (10) | (11) | (12) | (13) | (14) | (15) |
| #COVID / #ALL (Log10) | -0.583*** (0.148) | | | -0.419** (0.165) | | -0.665*** (0.163) | | | -0.425** (0.162) | | -0.320** (0.139) | | | -0.265* (0.141) | |
| (#COVID +1) / (#ALL +1) (Log10) | | -0.551*** (0.111) | | | -0.390*** (0.106) | | -0.561*** (0.106) | | | -0.414*** (0.106) | | -0.270** (0.108) | | | -0.250** (0.106) |
| Ave. Stringency | | | -0.580*** (0.108) | -0.320* (0.165) | -0.436*** (0.105) | | | -0.562*** (0.115) | -0.433*** (0.120) | -0.385*** (0.112) | | | -0.263** (0.126) | -0.206 (0.127) | -0.235* (0.122) |
| GDPpc (Log10) | | | | | | | | | | | 0.202 (0.125) | 0.166 (0.127) | 0.253** (0.124) | 0.206 (0.123) | 0.166 (0.124) |
| Population (Log10) | | | | | | | | | | | 0.489*** (0.115) | 0.455*** (0.118) | 0.528*** (0.113) | 0.480*** (0.114) | 0.439*** (0.115) |
| Day | | | | | | | | | | | -0.31*** (0.110) | -0.272** (0.113) | -0.197 (0.134) | -0.191 (0.131) | -0.133 (0.131) |
| $D^{COVID}$ | | | | | | 0.235 (0.163) | -0.260** (0.106) | -0.057 (0.115) | 0.205 (0.148) | -0.130 (0.104) | 0.146 (0.128) | -0.115 (0.099) | -0.012 (0.098) | 0.141 (0.126) | -0.078 (0.098) |
| Constant | 0 (0.148) | 0 (0.111) | 0 (0.108) | 0 (0.142) | 0 (0.097) | 0 (0.114) | 0 (0.106) | 0 (0.109) | 0 (0.103) | 0 (0.097) | 0 (0.088) | 0 (0.088) | 0 (0.089) | 0 (0.087) | 0 (0.086) |
| Observations | 32 | 59 | 59 | 32 | 59 | 59 | 59 | 59 | 59 | 59 | 59 | 59 | 59 | 59 | 59 |
| Adj. R2 | 0.318 | 0.292 | 0.325 | 0.376 | 0.449 | 0.248 | 0.349 | 0.316 | 0.381 | 0.455 | 0.546 | 0.554 | 0.539 | 0.560 | 0.576 |
| RMSE | 0.826 | 0.842 | 0.822 | 0.790 | 0.742 | 0.867 | 0.807 | 0.827 | 0.787 | 0.739 | 0.674 | 0.668 | 0.679 | 0.663 | 0.651 |
| $F$ Statistic | 15.46 | 24.89 | 28.90 | 10.32 | 24.65 | 10.56 | 16.56 | 14.38 | 12.91 | 17.11 | 14.97 | 15.41 | 14.56 | 13.29 | 14.13 |

Notes: The dummy variable $D^{COVID} = 1$ if $\#COVID = 0$, and $D^{COVID} = 0$ if otherwise. Countries where $\#COVID = 0$ are dropped in columns (1) and (4), and these countries are included in other columns after resetting the value of $\log_{10}(\#COVID/\#ALL)$ to 0. Standard errors in parentheses. Significant level: *p<0.1; **p<0.05; ***p<0.01.

In columns (1-3) of Table S8, we include alternative functional forms for the share of COVID-19 policy documents and the average stringency index, finding that the COVID share and the stringency index have independent and comparable predictive power for total deaths. In columns (4-5), we include both the policy document share and the stringency index and find that they augment each other's predictive power and have similar coefficient magnitudes. In columns (6-10), we include the dummy variable whose value $D^{COVID} = 1$ if a country did not publish any COVID-19 policy documents in the prior 30 days and $D^{COVID} = 0$ if otherwise. We find that the predictive powers of COVID share and stringency index are again similar to each other and augmentative.

In columns (11-15), we further include controls for GDP per capita, population, and the calendar day where 3 daily deaths were first recorded in each country. After controlling for all variables, we find that the effects of the share of COVID-19 policy documents remain negative and significant in columns (11-12). Moreover, the COVID share exhibits a slightly stronger power than the stringency index in predicting total deaths in columns (14-15). These observations are encouraging, as the stringency index measures policies that are specifically designed toward containing the pandemic, which exhibits a comparable level of predictive power as our measure. Overall these results further confirm the robustness of our results, showing that at the country level, greater COVID-19 policy attention predicts fewer subsequent deaths.

## SI Figures

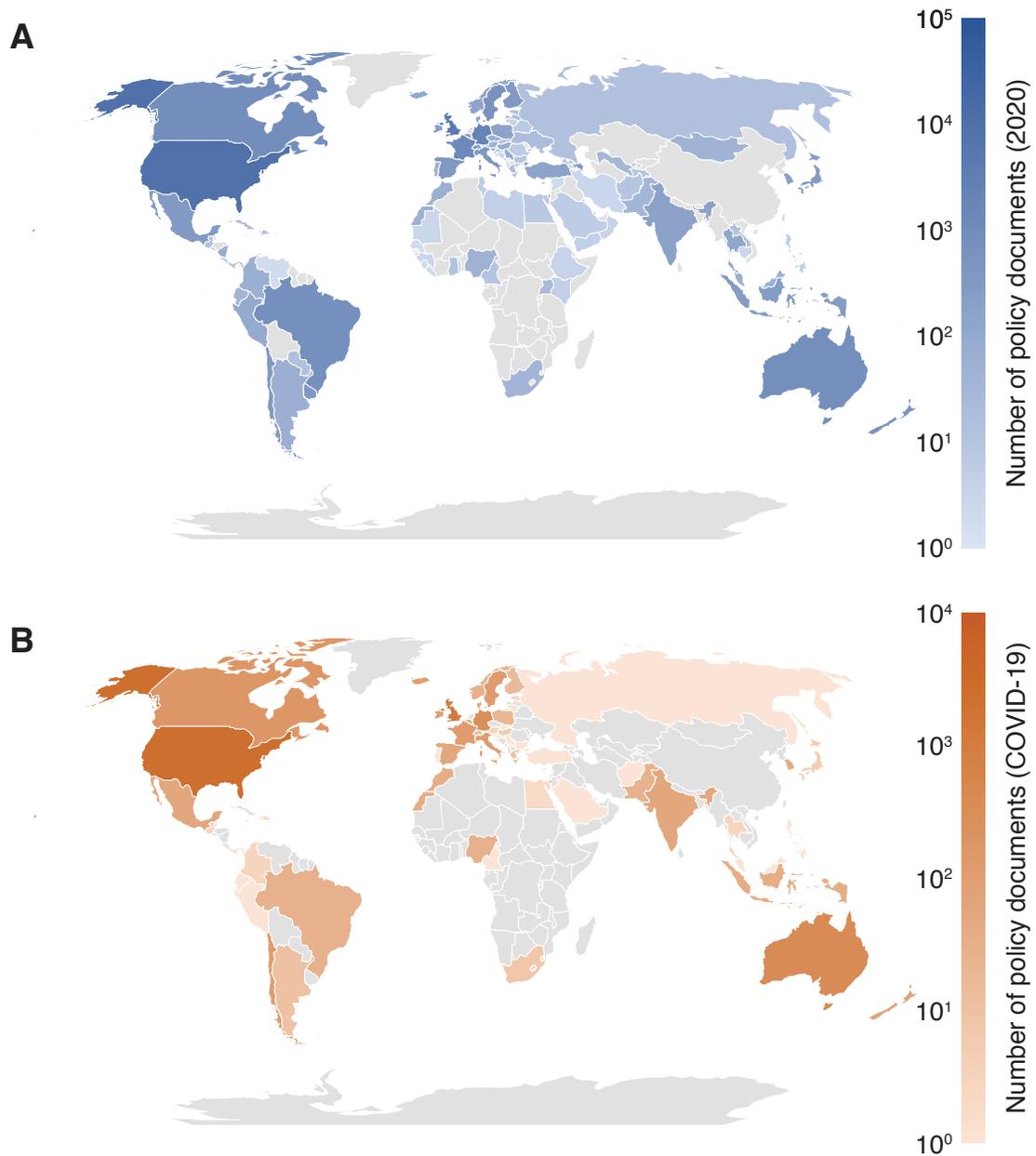

**Fig. S1 World map of policy data coverage. (A)** Number of policy documents published in 2020 in our data. **(B)** Number of COVID-19 policy documents in our regression analysis, where we restrict that a country has 3 daily deaths between January 31 and April 26, 2020; and has at least one policy document published in the 30 days before 3 daily deaths were first recorded. We visualize (#docs+1) for the logarithmic color scale.

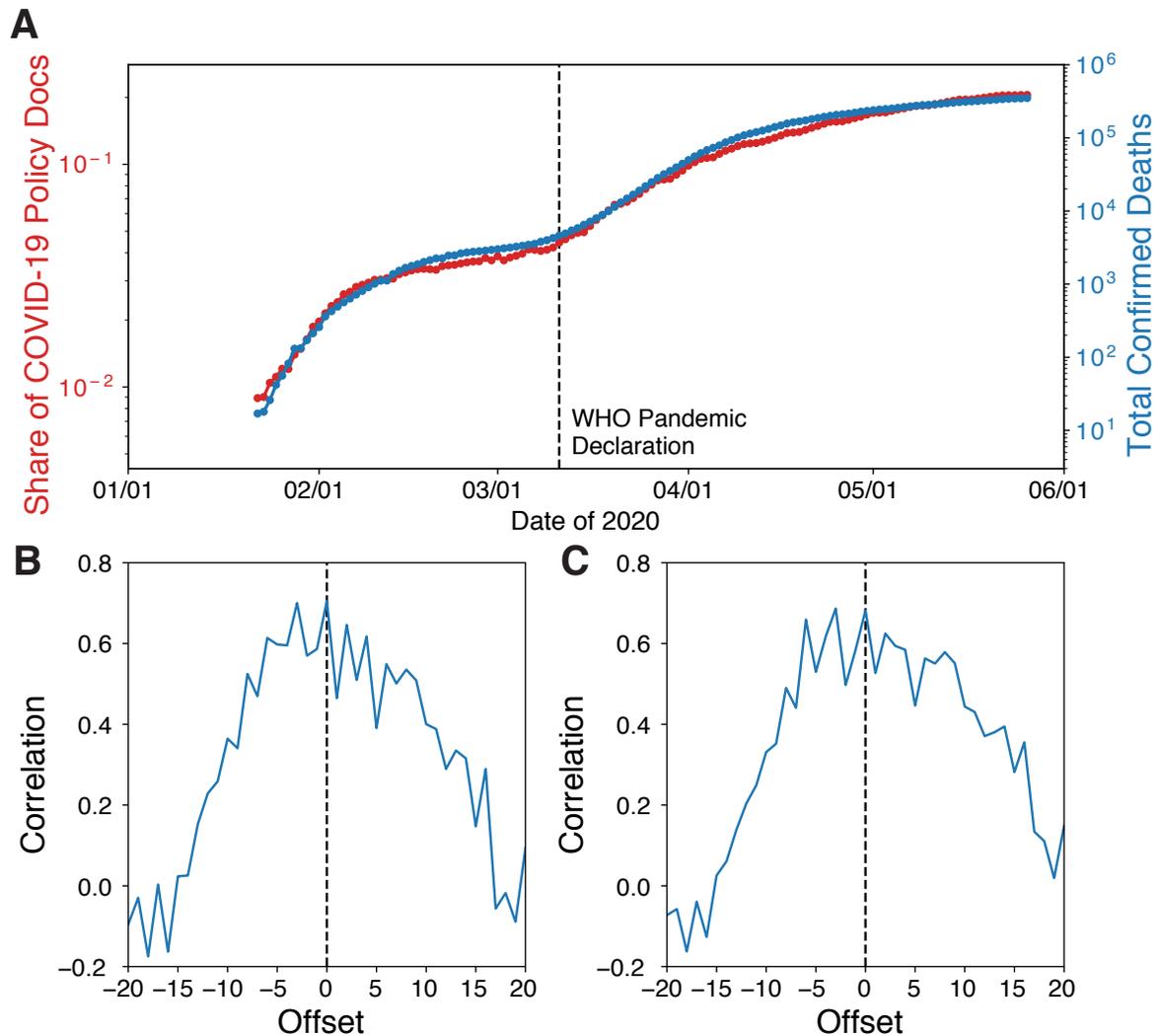

**Figure S2. Share of COVID-19 policy docs and the pandemic response. (A)** Trajectories of the global share of COVID-19 policy documents among all policy documents and the number of total confirmed deaths. **(B)** Pearson correlation between the shifted COVID-19 share curve and the total cases curve. **(C)** Pearson correlation between the shifted COVID-19 share curve and the total deaths curve. The offset is positive when shifting the COVID-19 share curve forward.

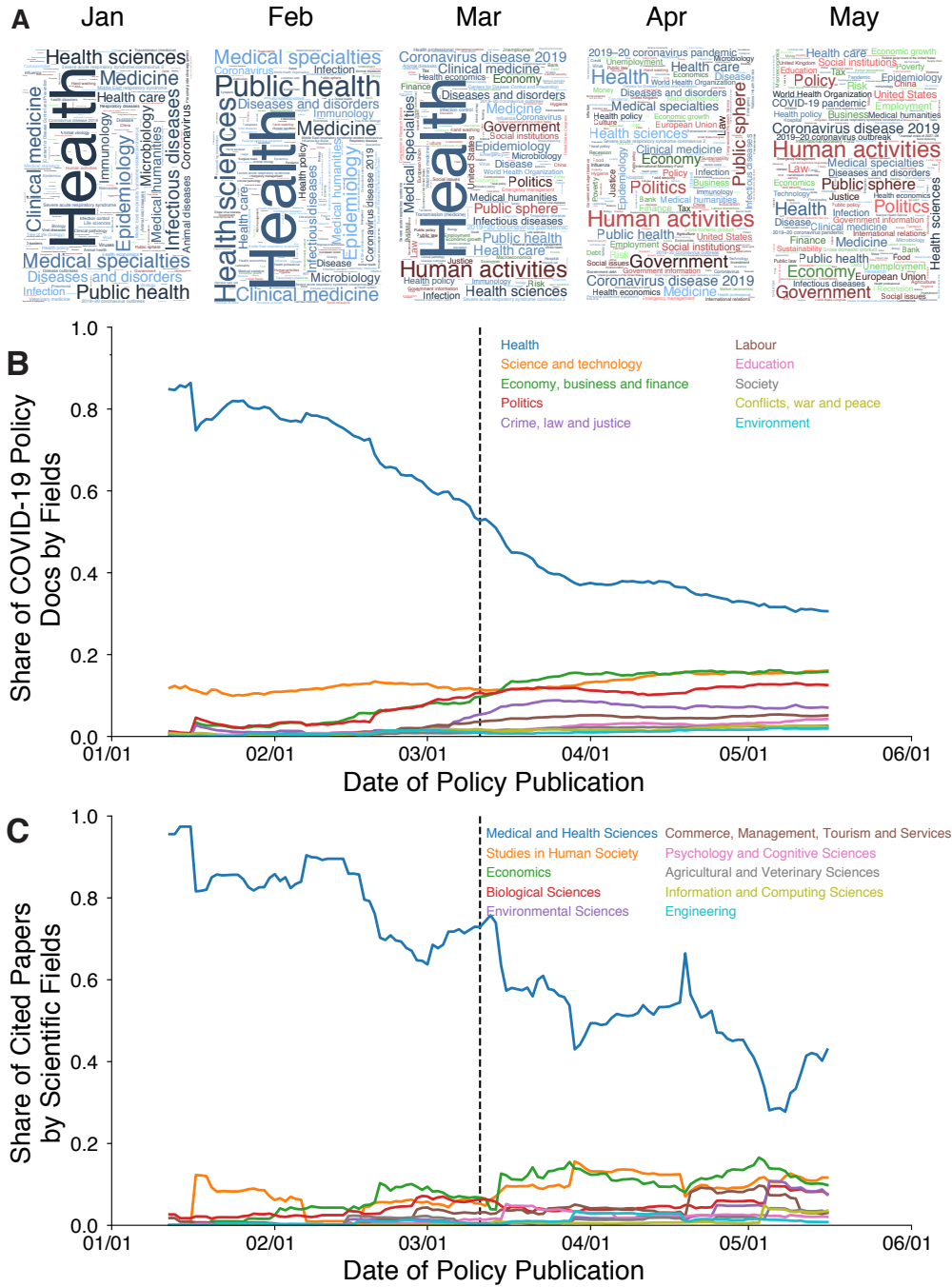

**Figure S3**. **Field evolution of COVID-19 policy documents and their cited papers. (A)** Word cloud of topics of COVID-19 policy documents published in each month of 2020. The size of a topic corresponds to the share of the topic among all topics in the documents. **(B)** The share of total COVID-19 policy documents across time by fields (21-day moving average). **(C)** Field share of total COVID policy cited papers across time by research fields (21-day moving average). Only the top 10 fields are presented.

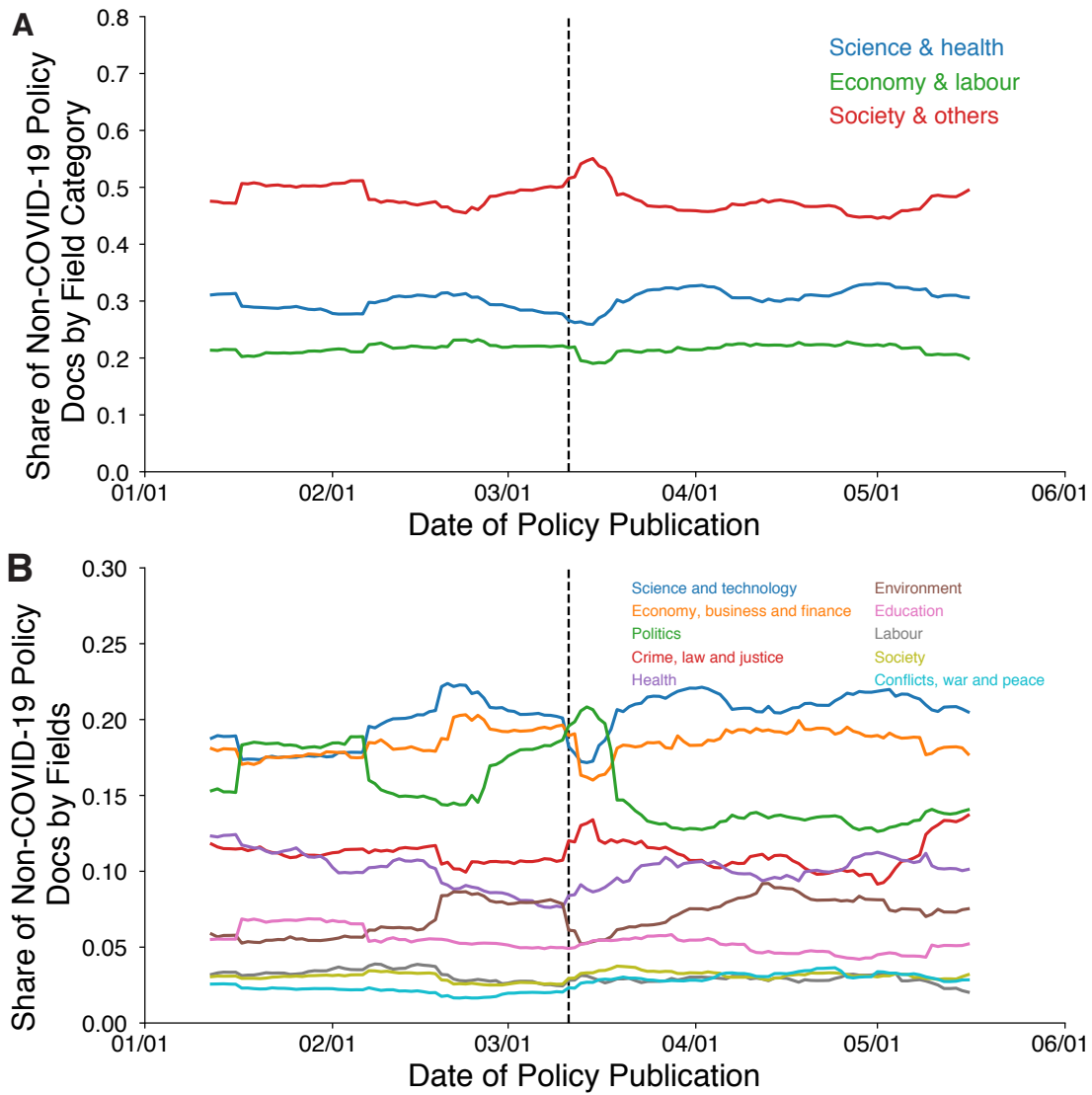

**Figure S4. Topic and field shifts of all other policy docs. (A)** The share of total non-COVID-19 policy documents across time by three field categories (21-day moving average). **(B)** The share of total non-COVID-19 policy documents across time by fields (21-day moving average). Only the top 10 fields are presented.

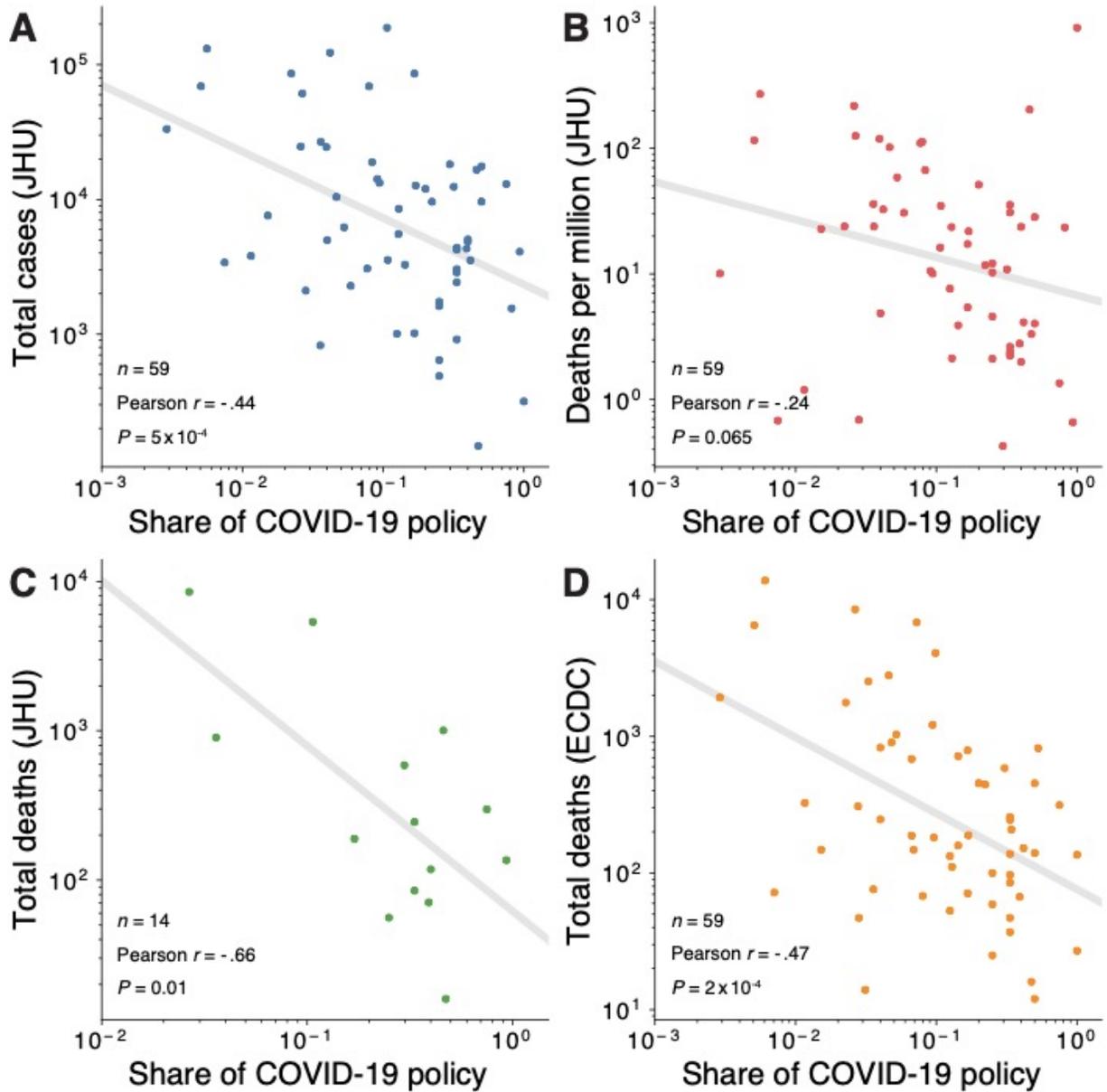

**Figure S5**. **The negative correlations between the share of COVID-19 policy attention and alternative measures of pandemic situation.** **(A)** The total confirmed cases recorded by Johns Hopkins University (JHU). **(B)** The deaths per 1 million population of a country recorded by JHU. **(C)** Results on total confirmed deaths recorded by JHU for countries where English is an official language. **(D)** The total confirmed deaths recorded by European Centre for Disease Prevention and Control (ECDC).

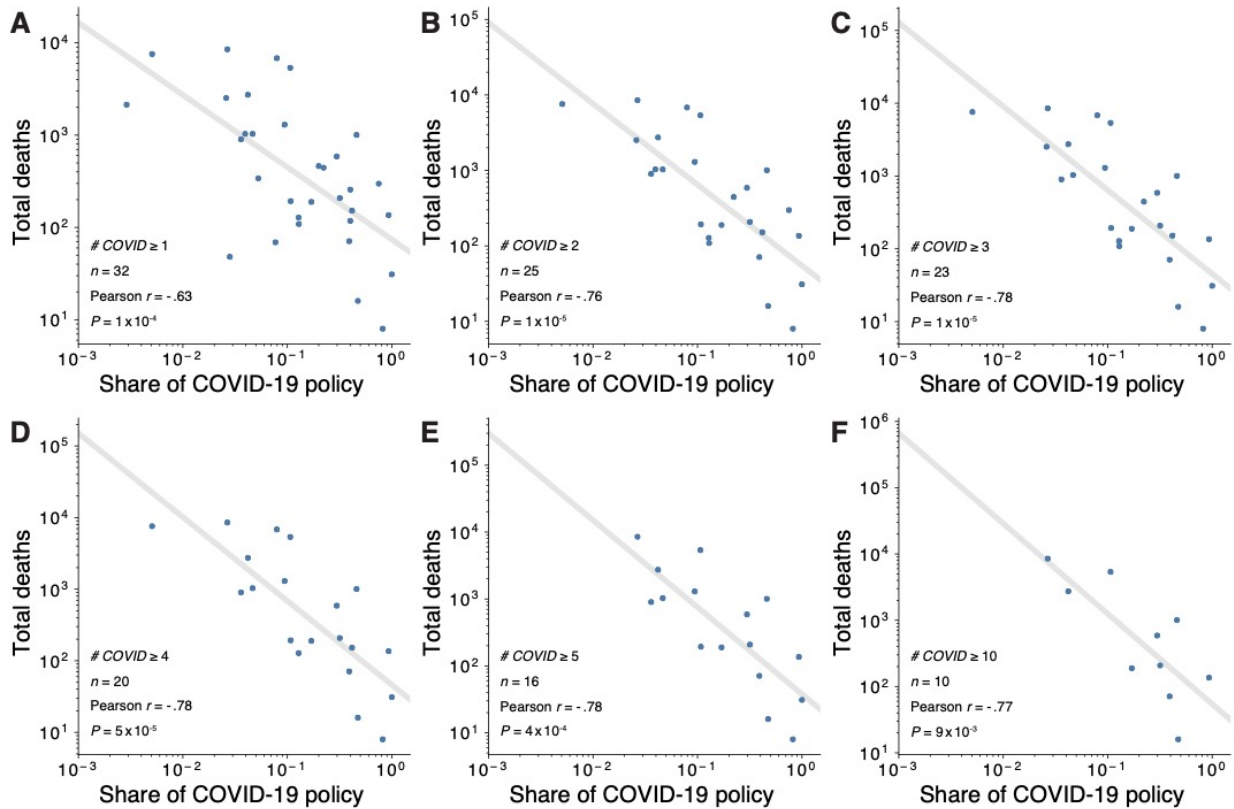

**Figure S6**. **Thresholds of the minimum number of COVID-19 policy documents.** The minimum number of COVID-19 policy documents is increased from 1 in panel **(A)** to 10 in panel **(F)**, as indicated in each panel. The negative correlation between COVID-19 policy attention and the pandemic situation tends to become steeper as the minimum number of COVID-19 policy documents increases.

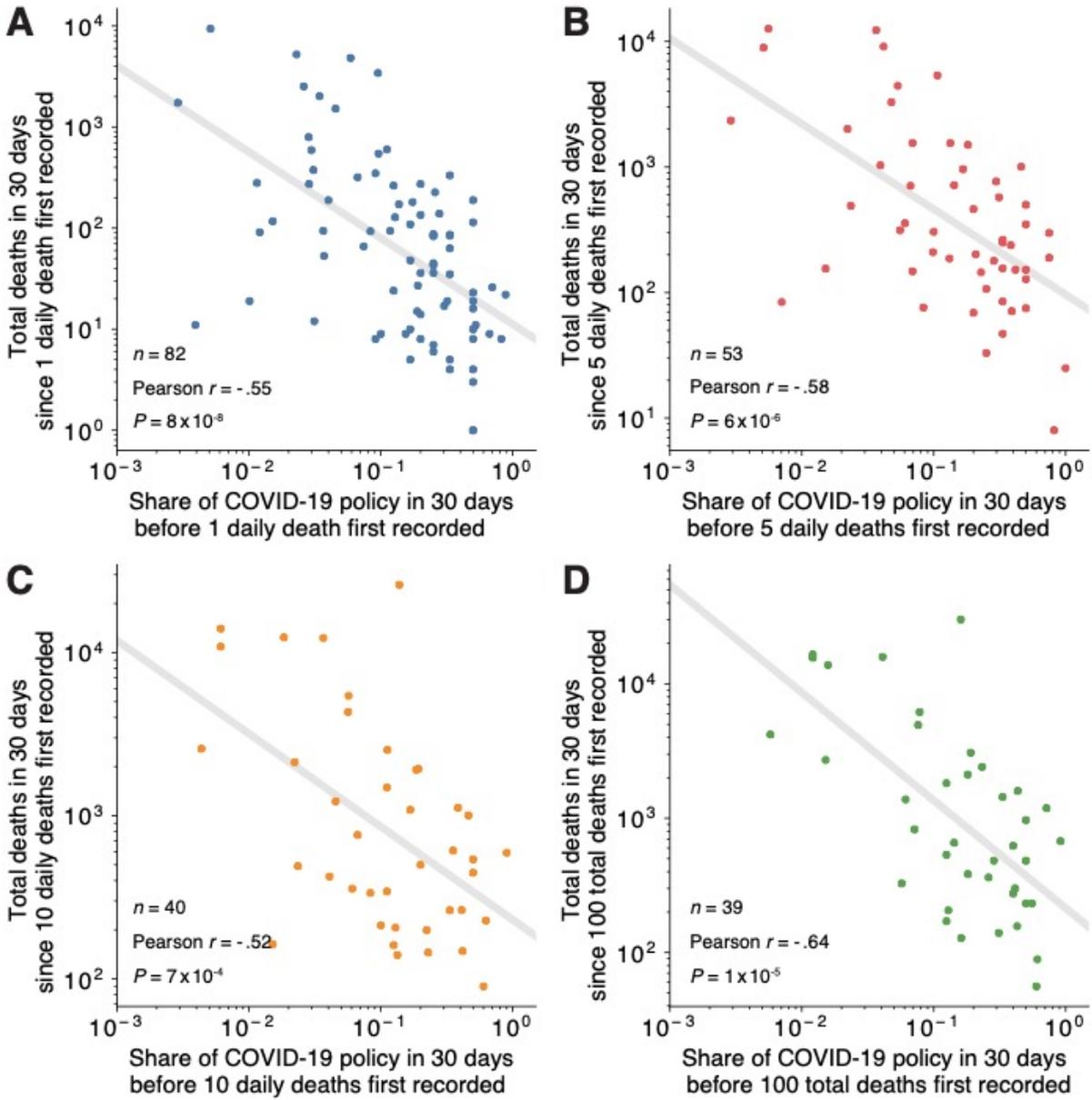

**Figure S7**. **Reference point timing for COVID-19 situation within each country. (A)** Reference point where the first death was recorded in each country. **(B)** Reference point where 5 deaths in a day were first recorded in each country. **(C)** Reference point where 10 deaths in a day were first recorded in each country. **(D)** Reference point where a total number of 100 deaths were first recorded in each country.

# References

1. Overton. *Overton Help Center: Advice and answers from the Overton Team*. 2020; Available from: http://help.overton.io/en/.

2. Adie, E. *Who are COVID-19 guidelines & policy documents citing?* 2020; Available from: https://blog.overton.io/blog/2020/04/23/who-are-the-covid-19-guidelines-policy-citing/.

3. Adie, E., *Personal communication.* 2020.

4. Herzog, C., D. Hook, and S. Konkiel, *Dimensions: Bringing down barriers between scientometricians and data.* Quantitative Science Studies, 2020. **1**(1): p. 387-395.

5. Dimensions Resources. *Dimensions COVID-19 publications, datasets and clinical trials*. 2020; Available from: https://dimensions.figshare.com/articles/Dimensions_COVID-19_publications_datasets_and_clinical_trials/11961063.

6. Dong, E., H. Du, and L. Gardner, *An interactive web-based dashboard to track COVID-19 in real time.* Lancet Infectious Diseases, 2020. **20**: p. 533-534.

7. European Centre for Disease Prevention and Control. *COVID-19 situation update worldwide*. 2020; Available from: https://www.ecdc.europa.eu/en/geographical-distribution-2019-ncov-cases.

8. Adie, E., *Who are COVID-19 guidelines & policy documents citing?* 2020.

9. Wikidata contributors. *COVID-19.* 2020; Available from: https://www.wikidata.org/wiki/Q84263196.

10. Albert, R., H. Jeong, and A.-L. Barabási, *Error and attack tolerance of complex networks.* Nature, 2000. **406**(6794): p. 378-382.

11. Schwartz, N., et al., *Percolation in directed scale-free networks.* Physical Review E, 2002. **66**(1): p. 015104.

12. Carmi, S., et al., *A model of Internet topology using k-shell decomposition.* Proceedings of the National Academy of Sciences, U.S.A., 2007. **104**(27): p. 11150-11154.

13. Kitsak, M., et al., *Identification of influential spreaders in complex networks.* Nature Physics, 2010. **6**(11): p. 888-893.

14. Financial Times. *Coronavirus tracked: Has your country's epidemic peaked?* . 2020 2020-05-30]; Available from: https://ig.ft.com/coronavirus-chart.

15. Hale, T., et al., *Variation in government responses to COVID-19*, in *Blavatnik School of Government Working Paper*. 2020.

16. Gibney, E., *Whose coronavirus strategy worked best? Scientists hunt most effective policies.* Nature, 2020. **581**(7806): p. 15-16.

17. Wikipedia. *List of territorial entities where English is an official language.* 2020; Available from: https://en.wikipedia.org/wiki/List_of_territorial_entities_where_English_is_an_official_language.